\documentclass[caption=false,font=normalsize,labelfont=sf,textfont=sf,lettersize, journal]{IEEEtran}

\usepackage{import}
\usepackage{packages}

\begin{document}
\title{Cell-free massive MIMO Channels in an Urban Environment - 
Measurements and Channel Statistics

\thanks{Part of this work has been submitted to IEEE VTC Fall 2024 \cite{zhang2024large}.}
\thanks{This work was financially supported partly by KDDI Research, Inc. and partly by the National Science Foundation. JGP's work was supported in part by the Foreign Fulbright Ecuador SENESCYT Program.}
\thanks{Yuning Zhang, Thomas Choi, Zihang Cheng, Jorge Gomez-Ponce, and Andreas F. Molisch are with the Wireless Devices and Systems Group in the Ming Hsieh Department of Electrical and Computer Engineering, \ac{USC}, Los Angeles, CA 90089, USA. Jorge Gomez-Ponce is also with the ESPOL Polytechnic University, Escuela Superior Politécnica del Litoral, ESPOL, Facultad de Ingenier\'ia en Electricidad y Computaci\'on, Km 30.5 vía Perimetral, EC090112, 
Guayaquil, Ecuador. Issei Kano and Masaaki Ito are with KDDI Research, Inc., Saitama, Japan. Corresponding author: Yuning Zhang (yzhang26@usc.edu).}
}

\author{}
\author{\IEEEauthorblockN{Yuning Zhang, Thomas Choi, Zihang Cheng, Jorge Gomez-Ponce, \IEEEmembership{Member,~IEEE}, Issei Kanno, Masaaki Ito,\\ and Andreas F. Molisch, \IEEEmembership{Fellow,~IEEE}}
}

\maketitle

\begin{abstract}
    \Ac{CF-mMIMO}, where each \ac{UE} is connected to multiple \acp{AP}, is emerging as an important component for 5G and 6G cellular systems. Accurate channel models based on measurements are required to optimize their design and deployment. This paper presents an extensive measurement campaign for \ac{CF-mMIMO} in an urban environment. A new ``virtual \ac{AP}'' technique measures channels between $\NumUETotal$ UE locations and more than $20,000$ possible microcellular \ac{AP} locations. Measurements are done at $3.5$ GHz carrier frequency with $350$ MHz \ac{BW}. The paper describes the measurement setup and data processing, shows sample results and their physical interpretation, and provides statistics for key quantities such as pathloss, shadowing, \ac{DS}, and delay window. We find pathloss coefficients of $\LOSPLAlphaOneDigit$ and $\NLOSPLAlphaOneDigit$ for \ac{LOS} and \ac{NLOS}, respectively, where the high \ac{LOS} coefficient is mainly because larger distance leads to more grazing angle of incidence and thus lower antenna gain in our setup. Shadowing standard deviations are $\LOSPLSOneDigit/\NLOSPLSOneDigit$ dB, and \ac{RMS} \acp{DS} of $\LOSDSMuOneDigit/\NLOSDSMuOneDigit$ dBs. The measurements can also be used for parameterizing a CUNEC-type model, which will be reported in future work.  
\end{abstract}

\begin{IEEEkeywords}
Cell-free Massive MIMO, Channel Sounding, Delay Dispersion, Delay Evolution, Urban Scenario
\end{IEEEkeywords}

\section{Introduction}
    The progression from fourth-generation (4G) cellular systems to 5G and beyond is motivated by the requirements to increase the data rates and the reliability and uniformity of data services \cite{tataria20216g}. An especially promising architecture to achieve these goals is \ac{CF-mMIMO}, which distributes in a geographical area a large number of \ac{BS} antennas in the form of single-antenna (or few-antennas) \acp{AP}, all of which are connected to a central processing unit via fast front haul connections \cite{zhang2019cell},\cite{demir2021foundations},\cite[Chapter 22]{molisch2023wireless}. Each \ac{UE} may connect wirelessly either to all \acp{AP} or a subset encompassing the \acp{AP} that are closest to it. This architecture provides more stable receive power than \acp{BS} with concentrated massive arrays since any \ac{UE} is close to at least one \ac{AP}. Furthermore, it eliminates inter-cell interference because there are no cells anymore.\footnote{\Ac{CF-mMIMO} is strongly related to the previously introduced concepts of base station cooperation, network MIMO, cooperative multipoint (CoMP), and Cloud-RAN.} Thus, a \ac{CF-mMIMO}  system outperforms small-cell systems in particular in terms of $95\%$-likely per-user throughput and offers more robustness compared to conventional cellular systems \cite{ngo2015cell, nayebi2015cell}.
    
    It is axiomatic that a good design of a wireless system must take into account the properties of the propagation channels in which the system will operate. The distributed nature of \ac{CF-mMIMO} systems requires, therefore, channel models that correctly describe the interrelations between the propagation channels {\em from multiple \acp{UE} to multiple \acp{AP}.} This is in contrast to the vast majority of current channel models (see also Sec. \ref{sec: state of the art}) that consider the link between one or more \acp{UE} to a {\em single} \ac{BS}. Obviously, such new models need to be based on new types of measurements.

    \subsection{State of the art}   \label{sec: state of the art}
       In contrast to the huge number of channel measurements performed with concentrated-MIMO setups, the number of {\em measurements} in \ac{CF-mMIMO} systems is very small. Measurements generally fall into two categories: (i) outdoor measurements, which are characterized by a small number of \acp{AP}, and (ii) indoor measurements, which often use virtual arrays for measurement and are easy to set up there. 

       In the former category, \cite{jungnickel2008capacity} focused on the measurement of capacity enhancement with four cooperating \acp{BS} that were all placed on walls surrounding a single courtyard; measurements were done at $5.2$ GHz. Ref. \cite{hammons2008cooperative} measured with a $3 \times 2$ system over a larger area, but in the $300$ MHz range, while \cite{maccartney2017base} measured the channel from \acp{UE} to two \acp{BS} at $73$ GHz. Another type of outdoor measurement involves the movement of the \ac{UE}. \cite{kurita2017outdoor} investigated the downlink between $3$ \acp{AP} and a moving \ac{UE} at $28$ GHz, with only \ac{LOS} channels. \cite{simon2023measurement} and \cite{loschenbrand2022towards} measure at lower frequencies, namely at $5.9$ GHz and $3.2$ GHz, with $8$ and up to $32$ \acp{AP}, respectively. 
       While these measurements were important for the early development of CoMP, they do not involve a sufficient number of \acp{AP} to allow modeling that is suitable for current \ac{CF-mMIMO} systems. Specifically, deployment planning and system optimization requires measuring at \emph{all potential} \ac{AP} locations instead of only at a few locations with inter-\ac{AP} distance separation, which in turn requires a much larger number of involved \acp{AP}. 

       In the indoor category, Ref. \cite{dey2019virtual} measured an $8 \times 8$ virtual setup at $2.53$ GHz in various rooms; \cite{choi2020co} measured with a $64 \times 64$ array in a single room at $3.5$ GHz, while \cite{tawa2020measuring} measured $8 \times 8$, but at $28$ GHz. \cite{zhang2013measurement} measured at 3.5 GHz with a $7 \times 7$ setup in various indoor environments, but only the \acp{AP} were distributed, while the 7 \ac{UE} antennas were concentrated on a single device. \cite{perez2021experimental} also measured at $3.5$ GHz, with $64$ different \ac{AP} locations in an indoor corridor environment. Both \cite{bultitude2017radio} and \cite{li2022toward} measured only in a single small room, but the former one measured at $2$, $18$, and $28$ GHz, and the latter reference measured at $2.6$ GHz with narrow \ac{BW}. While these measurements are interesting, they are by definition not applicable to the main deployment case of \ac{CF-mMIMO}, namely outdoor.  

       Large virtual outdoor setups were for the first time realized by the authors in \cite{choi2022using}, which suggested using a drone to fly a \ac{Tx} to a number of different possible \ac{AP} locations and measure there. The measurements achieved with this setup were used, e.g., in \cite{choi2021uplink} to assess the energy efficiency of \ac{CF-mMIMO} systems, and in \cite{choi2022realistic} to develop a new street-by-street pathloss model called CUNEC. However, these measurements were limited in two aspects: (i) the available control of the drone limited the precision of the locations at which the virtual \acp{AP} could be placed, and (ii) weight constraints required the use of a \ac{Tx} with relatively narrow \ac{BW} and poor phase stability.

    \subsection{Contribution of the paper}
        The current paper presents a measurement campaign that overcomes these limitations and provides the by far most extensive wideband measurements for \ac{CF-mMIMO} channels, using $\NumUETotal$ \ac{UE} locations and a total of more than $20,000$ \ac{AP} locations that are precisely tracked. The measurements are performed with $350$ MHz \ac{BW}. In particular, we
        \begin{itemize}
            \item describe a modification of our previous virtual array method that is better suited for wideband precision measurements. 
            \item present sample results of the measurements, together with the interpretation of the observed evolutions of impulse responses in terms of the environment geometry.
            \item provide the statistics of pathloss, shadowing, and \ac{DS} from this extensive measurement. 
        \end{itemize}

    \subsection{Paper organization}
        The rest of the paper is organized as follows: the measurement principle, the details of the channel sounder, and the measurement environment are described in Sec. \ref{sec: channel sounding system}. This is followed in Sec. III by a description of the data processing and extraction method for the desired channel parameters. Sec. \ref{sec: sample pdps} presents the sample results, while Sec. \ref{sec: parameter statistics} presents the statistics of pathloss, shadowing, and delay dispersion. A summary concludes the paper in Sec. VI.

\section{Channel Sounding System} \label{sec: channel sounding system}
    Our measurements were performed with a very large MIMO array, where the \ac{AP} side has a virtual array, created by mechanical movement of a single antenna along a trajectory, effected by a ``cherry-picker'' vehicle that raises the \ac{Tx} to a height of $13$ m, i.e., a typical \ac{AP} height. The \ac{UE} side consists of an $8$-element distributed switched array that remains static. The sounding of each \ac{SISO} link within this array is done through a wideband multi-tone waveform. The following sections describe the various components of this setup in detail. 

    \subsection{Channel Sounder}  \label{sec: channel soudner}
\begin{figure}[htp] 
            \centering
                \includegraphics[width=.9\linewidth, angle=0]{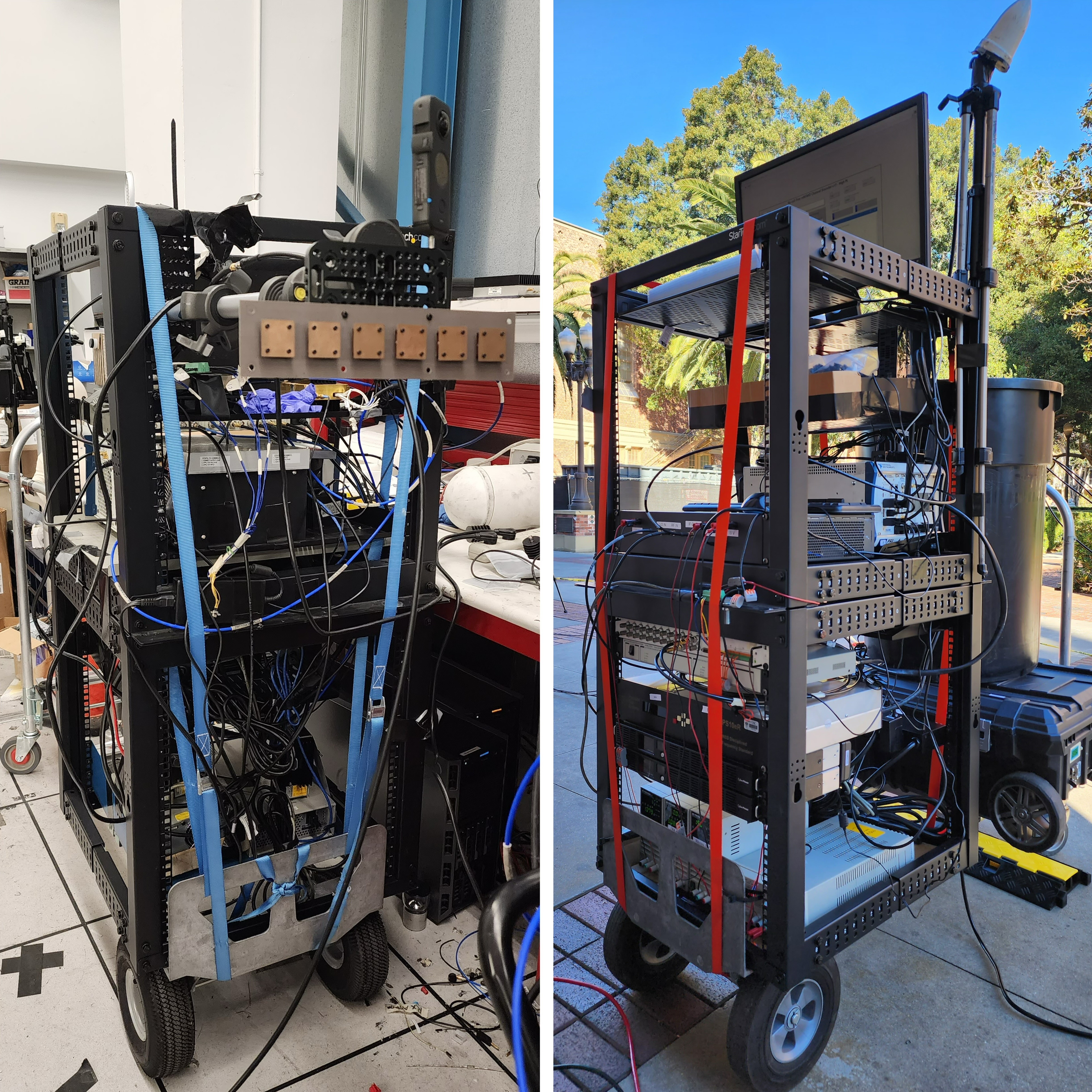}%
            \caption{ Photo of the Tx and the Rx.}
            \label{fig: system design}
        \end{figure}

        A photo of the complete setup can be seen in Fig. \ref{fig: system design}. Note that both the sounder equipment itself and the battery required to power it are much too heavy for mounting on a drone in the manner of the setup of \cite{choi2022using}. 

        \subsubsection{Transmitter}
\begin{figure}[htp] 
                \centering
                \subfloat[Azimuth cut at $90^\circ$ elevation]{%
                    \includegraphics[width=.9\linewidth, angle=0]{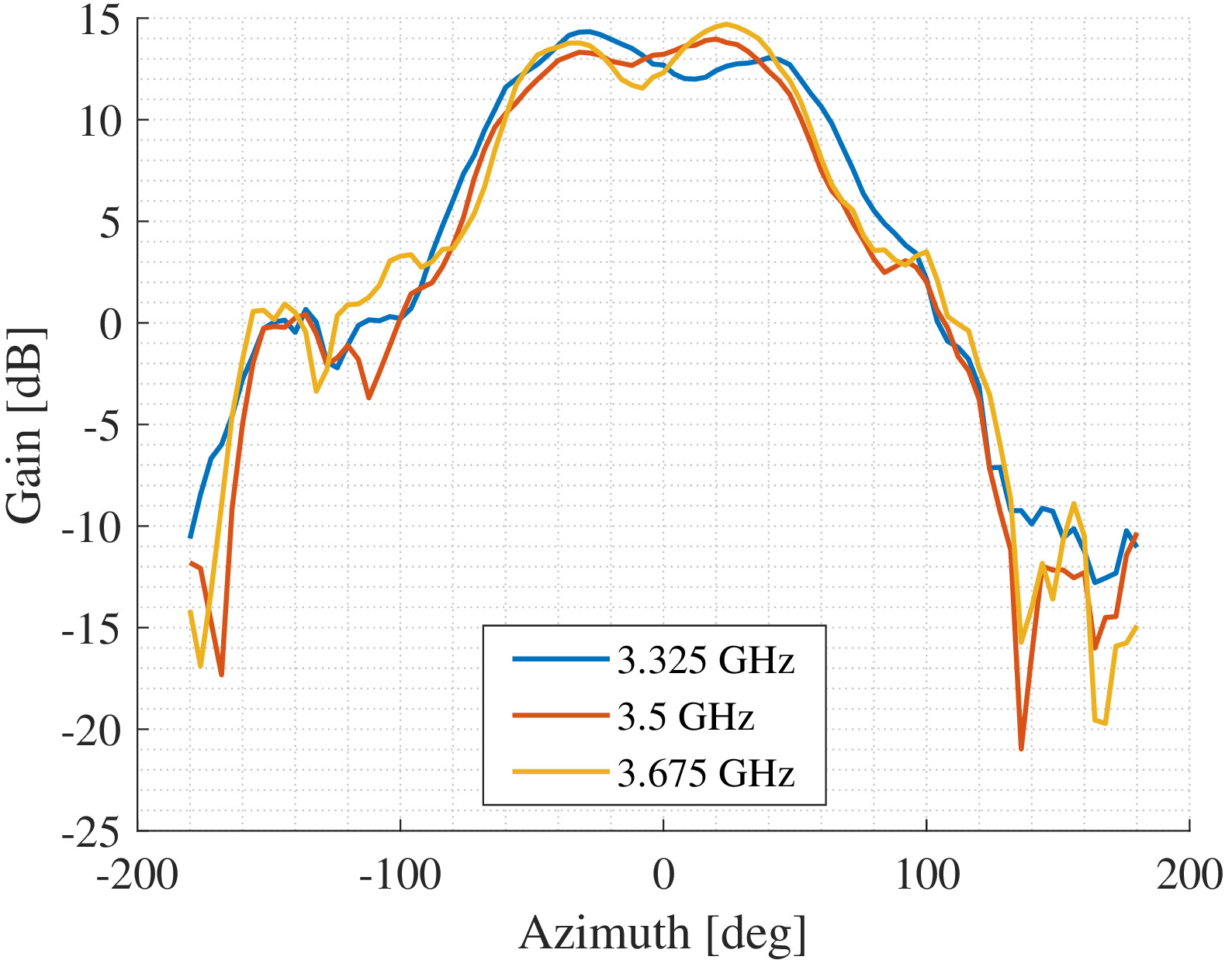}%
                    \label{fig: ula_pattern_az_cut}%
                    }\\
                \subfloat[Elevation cut at $0^\circ$ azimuth]{%
                    \includegraphics[width=.9\linewidth, angle=0]{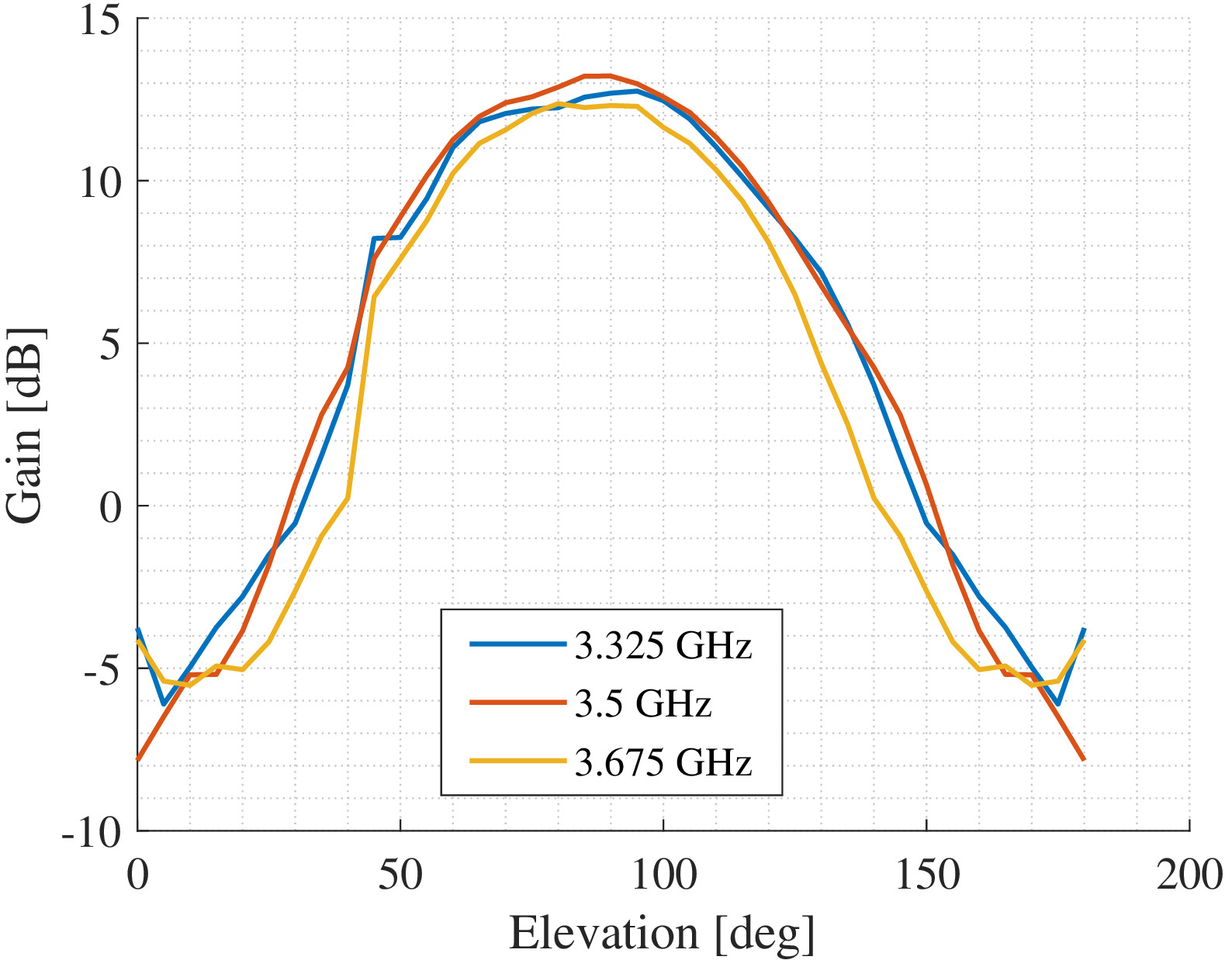}%
                    \label{fig: ula_pattern_el_cut}%
                    }%
                \caption{Azimuth and elevation cut of the pattern of the patch antenna.}
                \label{fig: ula_pattern_directional_cut}
            \end{figure}

             An \ac{AWG}, model Keysight N8241A, generates a sounding waveform with $350$ MHz \ac{BW} at an IF frequency of $375$ MHz. Such generation at digital IF has the advantage that no IQ mixers are required for upconversion, which would, by definition, be more prone to create distortions by IQ imbalances. This signal is mixed with a $3.875$ GHz sinusoidal oscillation from a precision frequency synthesizer (Eravant SOT-02220313200-SF-B6) to provide an RF signal with the center frequency of $3.5$ GHz. Two concatenated \acp{BPF} (Pasternack PE8713) are used to suppress the \ac{LO} leakage and the upper sideband, and send the signal to two concatenated \acp{PA} (Wenteq ABP1500-03-3730 and Mini-Circuits ZHL-100W-382+, with amplification of $37$ dB and $47$ dB, respectively), providing an output power level of $42$ dBm. The output is transmitted from a patch antenna element with vertical polarization (azimuth and elevation cuts, measured in our anechoic chamber with a standard gain horn, are shown in Fig. \ref{fig: ula_pattern_directional_cut}). A patch antenna was chosen as it can be expected to be the most prevalent antenna for \ac{CF-mMIMO} \acp{AP}, which are typically mounted on house walls.\footnote{Actually, the output of the \ac{PA} is sent to a $1$-to-$8$ high-power rated electronically-controlled RF switch follows after the \acp{PA}, which outputs to a \ac{ULA} that contains four dual-ports (horizontally and vertically polarized, respectively) antenna elements and two dummy elements on each end. An NI controller yields a timed sequence to the switch. The four active antenna elements are switched one after the other, and the horizontal port is always the first one to be selected for each antenna element. Note, however, that in the current measurement campaign, only the signal from the first (vertically) polarized antenna element was used. } During the measurements, the array was mechanically down-tilted at an angle of $40-50^\circ$, depending on the specific site location, to better align the main (vertical) transmission direction with the direction under which the UE is seen. 
        
            As a sounding sequence, we use a multi-carrier signal; it is similar to a Zadoff-Chu sequence used in LTE and 5G \ac{NR}, but maintains a lower \ac{PAPR} after filtering and over-sampling \cite{boyd1986multitone, conrat2006a, wang2017a, bas2019realtime}. Within the $350$ MHz \ac{BW} (resulting in a delay resolution of $2.857$ ns ($0.857$ m)), the waveform has $N_f=2801$ subcarriers, resulting in a subcarrier spacing of $\Delta_f=125$ kHz, corresponding to a maximum non-aliasing excess delay of $8\ \mu$s ($2400$ m). The waveform has a duration of $T_{\rm wfm}=8\ \mu$s. Bursts of $10$ repetitions of this waveform were transmitted for each location, which allows for improvement of the \ac{SNR} in post-processing (see Sec. \ref{sec: parameter processing}). 

        \subsubsection{Receiver}
            The \ac{Rx} antennas are 8 dipole antennas, with a nominally omni-directional pattern in the horizontal plane. Measurements of each of the patterns in our chamber showed variations with a standard deviation of $1.6$ dB in azimuth; we consequently marked the ``gain maximum'' direction for each antenna and ensured in our deployment that they always pointed in the same direction along the street, to obtain uniformity/reproducibility of the gain.\footnote{While precision dipoles with guaranteed uniformity are commercially available and would have been preferable, obtaining $8$ of them was cost-prohibitive.} Each of the antennas is connected to a \ac{LNA} (Mini-Circuits ZX60-53LN+ with amplification of $20$ dB and noise figure of $1.5$ dB) to compensate for the attenuation of the subsequent long ($30$ m) RF cables. The long cables enable distributing the \acp{UE} over a $60$ m large area, which is a requirement for our desired measurements (see also Sec. \ref{sec: measurement environment}). The cables are connected to an $8$-to-$1$ switch (Pulsar SP8T, operating frequency range up to $12$ GHz), whose output connects to a customized \ac{AGC} unit. This unit consists of two digitally controllable attenuators, one with attenuation switchable between $0$ and $30$ dB and one with continuously adjustable attenuation between $0$ and $120$ dB with a resolution of $0.05$ dB). Between these attenuators, two amplifiers of a total $59$ dB amplification are placed. This arrangement keeps the power variations of its output within $1.2$ dB standard deviation,\footnote{Strictly speaking, the AGC keeps the maximum of the power in the $8$ \ac{Rx} branches approximately constant, while the power of the other antennas can fluctuate more. The standard deviation for the strongest antenna is $1.2$ dB over the part of the trajectory with \ac{LOS} or strong \ac{NLOS}, and $4.5$ dB over the whole trajectory.} though the noise figure of the total setup can vary by up to $20$ dB (the high noise figure occurs in the case of high received power so that the minimum (over the different locations) of the dynamic range is not limited by this). The output of the \ac{AGC} is then mixed in a Mini-Circuits ZEM-4300MH+ mixer with an \ac{LO} (same as for the \ac{Tx}) for down-conversion to \ac{IF}. This signal is then digitized by a National Instruments digitizer, PXIe-5162. 

        \subsubsection{Synchronization and location recording}
            The system requires accurate synchronization for both switching control, carrier frequency, and sampling processes. This is achieved via two GPS-disciplined \ac{Rb} clocks (Jackson Labs HD CSAC $10$ MHz at the \ac{Tx} and Precision Test Systems GPS10eR at the \ac{Rx}). To achieve best accuracy, those clocks need to be trained beforehand; we did this by a combination of $12$-hour training inside our lab (where a GPS signal from a rooftop antenna is fed, via a cable, to the clocks), followed by another one-hour training when the equipment is outdoors to maintain the GPS tracking status and correct for inaccuracies arising during the short time between the clocks being disconnected from the GPS-carrying cables and the time the clocks are brought to an outdoor location where they have sufficient direct view of GPS satellites. 
            
            The clocks provide a $10$ MHz reference and a 1 \ac{PPS} signal. The $10$ MHz is used as a reference frequency/timing for all equipment, and the 1 \ac{PPS} is aligned to UTC and is the start trigger for both the switches and the transceivers, i.e., the \ac{AWG} and the digitizer.\footnote{As mentioned in Sec. \ref{sec: channel soudner}, the \ac{Tx} switched between different antenna elements, but only the signal from the first \ac{Tx} antenna element is used.}  The \ac{Tx} switches every $80\ \mu$s after a SISO is captured, corresponding to $10$ repetitions of waveforms. 

            The \ac{Rx} switches every $640\ \mu$s, thus requiring $5.12$ ms for a full \ac{Rx} switching cycle (only the first $80 \mu$s provide useful signal, as discussed in footnote 2). The system then goes idle while transferring the MIMO snapshot data from the host memory to an external hard drive. A full \ac{Rx} switching cycle occurs every $100$ ms. 

            Since measurements are done with a virtual array, possible environment changes during the measurements need to be observed (and, if necessary, corresponding measurement points must be omitted). Thus, we mounted $360^\circ$ cameras at each sounder and recorded at both \ac{Tx} and \ac{Rx} throughout the entire measurement campaign. Caution tapes are also used to prevent pedestrians from getting too close to the \ac{Rx} antennas and blocking them. 

            The \ac{Tx} was mounted in the basket of a cherry picker lift that is capable of driving with a raised basket; for safety reasons, a professional driver was used to operate the device. The driving speed was $0.4-0.6$ m/s, resulting in a spacing of the virtual array elements of $0.04-0.06$ m. Table \ref{table: meas_para} summarizes the most important measurement parameters.

            We stress again that while the creation of virtual arrays by \acp{UE} on the ground is a common technique for channel sounding, creating such a large, dense array of \acp{AP} has, to our knowledge, not been done before. 
            
            \begin{table}[h]
                \centering
                \caption{Measurement parameters}
                \begin{tabular}{|c|c|}
                    \hline
                    \textbf{Parameter} & \textbf{Value} \\ \hline \hline
                    AP movement speed & $0.4-0.6$ m/s \\ \hline
                    AP/UE height & $13$ m / $1$ m \\ \hline
                    AP/UE closest horizontal distance & $3$ m \\ \hline
                    AP antenna element $3$ dB beamwidth & $100^\circ$ \\ \hline
                    Post-processing SNR gain & $10$ dB \\ \hline
                    SISO duration & $80$ $\mu$s \\ \hline
                    SIMO duration & $640$ $\mu$s \\ \hline
                    MIMO duration & $5.12$ ms \\ \hline
                    MIMO burst rate & $10$ Hz \\
                    \hline
                \end{tabular}
                \label{table: meas_para}
            \end{table}

    \subsection{Calibration Procedure}
        Evaluation of the measurements requires calibration of the sounder. This occurs in two steps: (i) \ac{B2B} calibration, and (ii) antenna calibration. 
        For the former, the output of the \ac{Tx} switch is connected not to the \ac{ULA}, but rather, via a suitable attenuator, to the \ac{Rx} input (note that the \ac{Rx} includes the cables and the switch). The \ac{Tx} then sends its sounding sequence, and the \ac{Rx} digitizes this signal, thus establishing the transfer function. Since both \ac{Tx} and \ac{Rx} have 8 antenna ports each, we have a set of $64$ \ac{B2B} calibrations.  

        In a separate step, we calibrated the frequency-dependent antenna gain characteristics of the \ac{Tx} antenna, as well as of all the \ac{Rx} antennas. This calibration was done in the anechoic chamber at \ac{USC}, with a precision rotor rotating the antenna in $4^{\circ}$  (in azimuth) and $5^{\circ}$  (in elevation) steps, while the other link end is a calibrated standard gain antenna, and a \ac{VNA} to measure at the different frequencies. For the \ac{Tx} antennas, we measured $360^{\circ}$ azimuth and $45-180 ^{\circ}$ elevation, and calibrated both by themselves and in combination with the switch; while for the \ac{Rx} antennas, $360^{\circ}$ azimuth, and $0-180 ^{\circ}$ elevation was measured.

    \subsection{Measurement Locations}   \label{sec: measurement environment}
\begin{figure}
            \centering
            \includegraphics[width=1\linewidth]{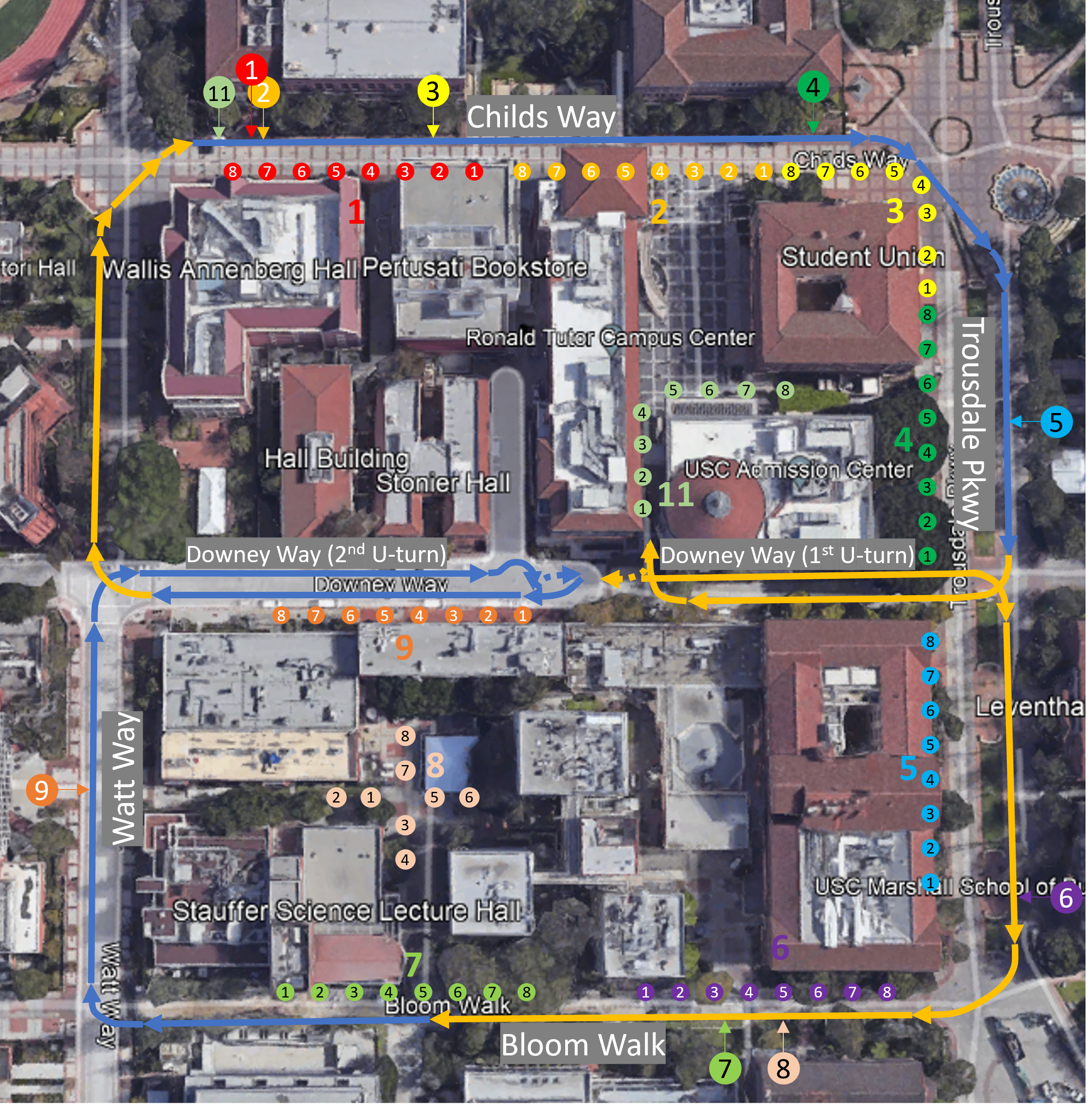}
            \caption{Small colored dots: measurement UE site locations, each color represents one UE cluster. Large colored circle with an arrow: start/stop location of \ac{AP} trajectory for a particular cluster.}
            \label{fig: measurement locations}
        \end{figure}

        Measurements were performed in an urban (but not metropolitan) area in downtown Los Angeles, CA, USA, namely the \ac{USC} University Park Campus. The measured area encompasses two city blocks, which cover a square of about $200 \times 200$ m, see Fig. \ref{fig: measurement locations}. It is characterized by medium-height buildings (typically $3\sim4$ floors) interspersed with alleys and plazas; the width of the streets is between $3$ and $15$ m, depending on the specific street location.
        
        The \ac{AP} moves on a trajectory that covers the outer streets once, where the AP drives on the ``further side'' (with respect to the encircled street blocks) of the street, with the \ac{Tx} antenna broadside pointing ``inwards'' perpendicular to the driving direction. The middle street is covered twice, with the \ac{Tx} on different sides of the street, and pointing into different directions.\footnote{The cherry picker changes directions in the middle of that street, going in reverse to change direction; this is indicated by the dashed part of the line in Fig. \ref{fig: measurement locations}.} The \ac{AP} maintains $13$ m height throughout the trajectory as a typical \ac{AP} height \cite{3GPPTR38901} (denoted as the blue line in Fig. \ref{fig: measurement locations}), except for certain locations where the AP has to be lowered to $4 \sim 5$ m to avoid collision with tree canopies (shown as the yellow line). The \acp{Rx} are placed in $10$ clusters of $8$ antenna each, where the antennas are distributed over a $60$ m wide area, such that the spacing between two adjacent \acp{UE} is approximately $8$ m. For each placement of a UE cluster, the cherry picker drives along the whole trajectory. Due to setup and measurement time, only 2-3 clusters could be measured each day, and the total measurements were extended (including break days) for two weeks. No significant changes in the environment (e.g., due to construction) occurred during this time.  
        
        The trajectories thus encompass both \ac{LOS} and \ac{NLOS} situations, where we define the latter as blockage of the \ac{LOS} by a building. Furthermore, parts of the \ac{LOS} portions of the route are obstructed by foliage, which we henceforth denote as \ac{OLOS}. The tree canopies typically start at $8$ m above the ground and extend beyond the height of the \ac{AP}. 

    \subsection{Clock drift correction}
\begin{figure}
            \centering
            \includegraphics[width=1\linewidth]{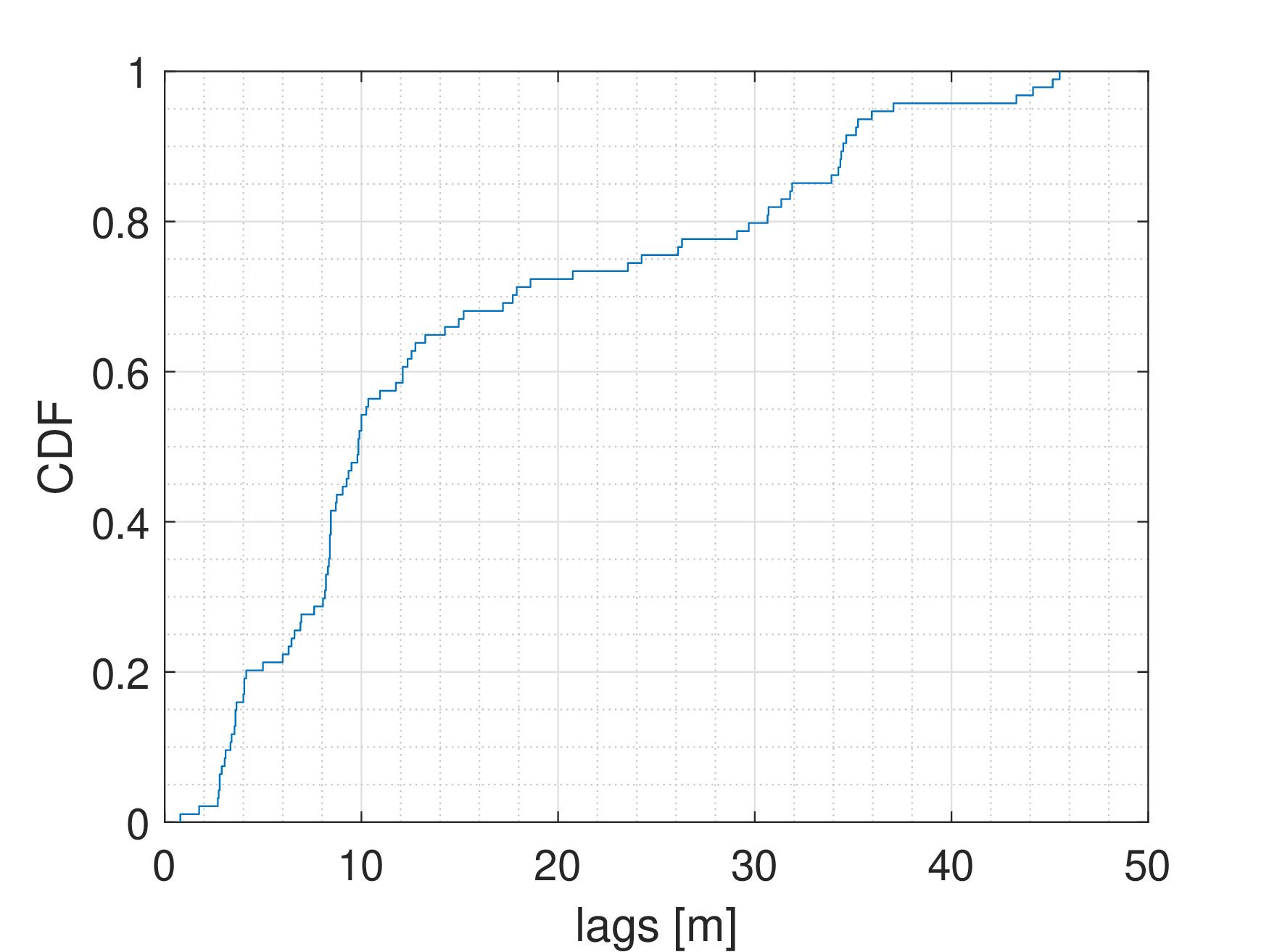}
            \caption{CDF of the clockdrift decorrelation distance from all LOS channels}
            \label{fig: gps_drift_uncorr_distance_cdf}
        \end{figure}

        The movement of the \ac{AP} may lose the locking of GPS timing signals, pushing the \ac{Rb} clocks at the \ac{Tx} and \ac{Rx} into free-running mode with increased phase drift. We performed a delay domain $2$-stage correction based on the channel status. For \ac{LOS} channels, we use the locations of \acp{AP} and \acp{UE} acquired from the synchronized camera recordings from the real-time video recordings to compute the \ac{AP}-\ac{UE} distance as the true reference to compute the delay correction offset; for \ac{NLOS} channels, we assume, 
         based on the \ac{LOS} measurements, that the phase drift is a linear function of time and compute the correction offset by linearly interpolating between the phases at the first/last LOS point next to a NLOS region. To estimate the residual error, we also perform a linear interpolation in the LOS region, and compare this interpolation against the true phase drift that we obtain as described above. The \ac{RMS} error has an expected value of $0.43$ m, and the standard deviation is $0.64$ m. The CDF of the correlation distance of the error (i.e., the distance between two points such that the correlation between their distance offsets is $\frac{1}{e}$) from different \ac{LOS} channels shows a mean value of $10$m and for the $10\%$ worst cases a correlation distance of $3$ m, see Fig. \ref{fig: gps_drift_uncorr_distance_cdf}. This indicates that the (unpredictable) deviations of the clock phase from the linear drift change relatively slowly. 

\section{Data Processing} \label{sec: parameter processing}
    This section describes the post-processing of the measured data to extract the channel impulse responses and condensed parameters derived from them. 

    \subsection{Pre-processing of data}
        The received data are the temporal samples at the \ac{Rx}, where each burst (\ac{Rx} cycling through the \ac{Rx} antennas, lasting for $5.12$ ms) is stored as a separate file named with the snapshot index (which can be related to the timestamps and thus locations in post-processing). We first perform a \ac{FFT} over the captured time-domain data; since each such file represents a $10$-times repetition of the sounding sequence, we then perform an averaging over those repetitions, resulting in a $10$ dB SNR gain. We divide the result by the \ac{FFT} of the sounding sequence, providing the ``raw'' transfer function of the channel $H_{{\rm raw},m,j,k}$, where $m$ is the snapshot (time) index, $j$ is the index of the \ac{Rx}, and $k$ is the frequency index. 

         From the digitizer (which has a sampling rate of 1.25 GS/s), we crop the $350$ MHz IF band corresponding to the desired $3.325-3.675$ GHz passband.  The cropped transfer function is then divided by the (cropped) B2B calibration transfer function $H_{{\rm cal},k}$. Note that the current measurements are not directionally resolved and all presented results contain the impact of both the channels and the antennas. Consequently, the transfer functions of both \ac{Tx} and \ac{Rx} antennas, which show some dependence on the considered direction (see, e.g., Fig. \ref{fig: ula_pattern_az_cut}),  are contained in the transfer functions. The calibrated transfer function is then filtered with a Kaiser-Bessel filter with parameter $\beta=3$ to reduce sidelobes in the delay domain. While the 3-dB \ac{BW} is reduced by about a factor of $2$ through this filtering, it is still larger than the typical band assignment for a 5G operator, i.e., $100$ MHz. The transfer function is then furthermore zero-padded (extending the \ac{BW} by a factor of $10$), resulting in oversampling of the impulse response (its Fourier transform) as described below; this is important for accurate computation of the \ac{DS} \cite{gomez2023impact}.    

        From the calibrated, cropped, filtered, zeroes-extended transfer function $H_{m,j,k}$, we then obtain the impulse response via \ac{IFFT}, and from that the \ac{PDP}:
        \begin{align}
            P_{{\rm h,}{m,j,\tau}} = \Big|\mathcal{F}^{-1}_{k} \big\{  H_{m,j,k}  \big\}\Big|^2    \label{eq: PDP}
        \end{align}
        where $\mathcal{F}^{-1}(\cdot)$ denotes the inverse Fourier transform with respect to $k$, $\tau$ is the index in the Fourier dual domain (i.e., delay), and $\mathcal{F}^{-1} \big\{  H_{m,j,i}(f)  \big\}$ represents the complex CIR.  
        We use here $\tau$ to denote any delay indexing scheme, irrespective of whether it is oversampled or not; the meaning is clear from the context.
        We then apply \ac{SSA} to all the \acp{PDP} using a sliding window with a width of $0.5$ m (corresponding to $6$ wavelengths). Note that this window length is a compromise between residual \ac{SSF} and the need not to obscure actual changes in the \ac{SSA} statistics; in particular, the delay-domain change of a \ac{MPC} within this window should be smaller than the resolvable delay bin. 
        
        We subsequently introduce thresholding at a level $\theta_{m',j}$ (on a dB scale) that is the maximum of a threshold $\theta_{{\rm n},m',j}$ that is $\Delta_{{\rm n}}$ (set to 7 dB) above the average noise level $P_{{\rm n},m',j}$ of this \ac{SSF} averaging window $m'$, and a threshold that is $\Delta_{{\rm dr}}$ (set to 20 dB) below the peak of the \ac{PDP} in this window: 
        \begin{align}
            \theta_{m',j} = \max \Big(P_{{\rm n},m',j}+ \Delta_{{\rm n}},\ \underset{\tau}{\max} \big(P_{{\rm h},\tau}\big)-\Delta_{{\rm dr}} \Big)   \label{eq: noise threshold level}
        \end{align}
        We furthermore apply delay gating to the impulse response, such that all delay bins corresponding to a distance of $>343$ m are set to zero. This was done since, in the considered geometry, no significant \acp{MPC} can be expected for longer delays. 
        The combined operation of noise thresholding and delay gating can be written as \cite{abbasi2023thz}
        \begin{align}
            P_{\rm h}^{\rm DG;NT}(\tau) = \Big[  P_{\rm h}(\tau): (\tau \leq \tau_{\rm gate}) \cap  (P_{\rm h}(\tau) \geq \theta_{m',j})  \Big]     \label{eq: DG & NT}
        \end{align}

        A further important processing step is the elimination of impulse response precursors. In our measurements, we observed significant energy arriving {\em before} the \ac{LOS} component (whose delay could be easily determined from the geometrical location of \ac{AP} and \ac{UE}). These precursors were not wraparound components (this could be excluded from the fact that such components would require a run length of more than $2$ km). Rather, they are direct ``talk through'' from the \ac{Tx} antenna into the \ac{Rx} device, without the detour via the \ac{Rx} antenna and cable. This explanation was confirmed by the fact that the precursors showed up at all \acp{UE} irrespective of their locations, and were furthermore exactly at the delay (which changed with \ac{Tx} location) corresponding to the distance between the \ac{Tx} antenna and the \ac{Rx} sounder box. After having confirmed the source of the precursors, we could thus simply eliminate them in post-processing,  i.e., all components arriving more than a ``guard interval'' (accommodating the Kaiser pulse corresponding to the LOS component) of 4 delay bins before the LOS component are set to zero. Note that this issue did not occur in NLOS situations, due to the large separation between \ac{Tx} and \ac{Rx} box.

    \subsection{Definition of parameters}
        \subsubsection{Pathloss and shadowing}
            The \ac{PG}, which is the inverse of the pathloss, is computed as  
            \begin{align}
                PG = \sum_\tau P_{\rm h}(\tau)    \label{eq: PL}
            \end{align}     
            For the \ac{PG} we furthermore perform an averaging over a window of size $20$ wavelength; such a larger window (compared to the window used for eliminating the SSF in the PDP) is permissible here because \acp{MPC} moving from one resolvable delay bin to another is not a concern anymore; it has furthermore the advantage of better averaging out residual \ac{SSF} fluctuations.    

            We will further use an $\alpha-\beta$ model to describe the pathloss as a function of distance between AP and UE:
            \begin{align}
                PL_{\rm dB}(d) = \alpha \cdot 10\log_{10}(d) + \beta  + S    \label{eq: a-b model}
            \end{align}
            where $\alpha$ and $\beta$ are the linear fitting coefficients, and $S$ is a random variable describing shadowing; it is commonly assumed to be zero-mean Gaussian random variable (i.e., lognormal on a linear scale), and is characterized by its shadowing distribution. We will parameterize this model from our measurements because it is widely used for system simulations of \ac{CF-mMIMO} \cite{demir2021foundations}. However, it is worth noting that it does not reproduce important effects in urban microcells. The CUNEC model recently introduced by us \cite{choi2022realistic} is better suited for such systems, but is more complex, so that its parameterization from our measurement campaign is left for future work.   
    
            We finally note that although the distance between \acp{AP} can be non-uniform due to acceleration/deceleration of the cherry picker, the density changes are small. Thus, there is no need to perform a {\em weighted} pathloss model fitting as proposed in \cite{karttunen2016distance, gomez2021air}. 

        \subsubsection{Delay dispersion parameters}
            The \ac{DS} is the most common condensed parameter for the description of the delay dispersion. It is defined as the second central moment of the \ac{PDP}
           \begin{align}
                \sigma_\tau = \sqrt{\frac{\sum_\tau P_{\rm h}(\tau)\tau^2}{\sum_\tau P_{\rm h}(\tau)}  -  \Bigg( \frac{\sum_\tau P_{\rm h}(\tau)\tau}{\sum_\tau P_{\rm h}(\tau)} \Bigg)^2} ~.    \label{eq: RMS delay spread}
            \end{align}
            The \ac{DS} depends significantly on the dynamic range. In order to avoid distortions of the statistics, we therefore only consider those \ac{PDP} windows that have $20$ dB dynamic range for the computation of the \ac{DS} statistics. 

            While the \ac{DS} is commonly used, it gives excessive weight to long-delayed \acp{MPC}. From a system design point of view, a better measure is the Q-window and the Q-tap number \cite[Chapter 6]{molisch2023wireless}. Defining a threshold \textcolor{black}{$\gamma= \frac{SIR_{\rm ISI}}{SIR_{\rm ISI}+1}$} (where  $SIR_{\rm ISI}$ is the desired signal-to-intersymbol-interference ratio) the Q-window is defined as
            \begin{equation}
                Q_{\rm win} = \underset{x}\min ~x,~ s.t. ~\underset{T_0}\max   \sum_{\tau=T_0}^{T_0+x} P_{\rm h}(\tau)\geq \gamma PG . 
            \end{equation}
            
            We furthermore define the Q-tap number, using the {\em sorted} power in the resolvable delay bins \textcolor{black}{$PDP_{\rm sort}(\tau_m)$} as:
            \begin{equation}
                  Q_{\rm tap} = \underset{x}\min ~x,~ s.t. ~ \sum_{m=1}^{x} PDP_{\rm sort}(\tau_m) \geq \gamma PG, 
            \end{equation}
            
            Q-window describes the length of an equalizer with contiguously-placed taps (or cyclic prefix in an OFDM system) that is necessary to achieve a certain $SIR_{\rm ISI}$. Q-tap describes the number of taps in an equalizer (or Rake fingers in a Rake receiver) that can place the taps at arbitrary delays.\footnote{By length of an equalizer, we mean the length of the impulse response that the equalizer has to compensate. This may be different from the actual number of taps used in the linear filter that constitutes the equalizer \cite{proakis2007}.}

\section{Sample PDPs} \label{sec: sample pdps}
    This section will describe sample measurement results and relate them to the geometry of the environment. It serves both to point out important propagation effects and as a sanity check for the measurements, supporting thus the total channel statistics that will be described in Sec. \ref{sec: parameter statistics}. We note, however, that only delay information, and no angular information is available. Thus, we can verify whether propagation paths that are plausible (due to the mechanisms such as reflection at large buildings) are consistent with the observed delays,\footnote{We further verified the potential propagation paths by comparing the evolution of their delay as to \ac{AP} is moved to/from the location of the plotted \ac{PDP} location, similar to the discussion in Sec. \ref{sec: AP-location-dependent PDP}; for space reason these comparisons are not discussed further here.} but it is not possible to reconstruct with certainty the involved paths. Instead of using $\theta_{m',j}$ for parameter computation, all \ac{PDP} examples presented in this section are processed with noise thresholding condition only, i.e., $P_{{\rm n},m',j}+ \Delta_{{\rm n}}$, to enable a richer presentation of propagation paths. 

    \subsection{Snapshot of PDP}
        \subsubsection{LoS PDP}
\begin{figure}[htp] 
                \centering
                \subfloat[Short LoS PDP]{%
                    \includegraphics[width=1\linewidth, angle=0]{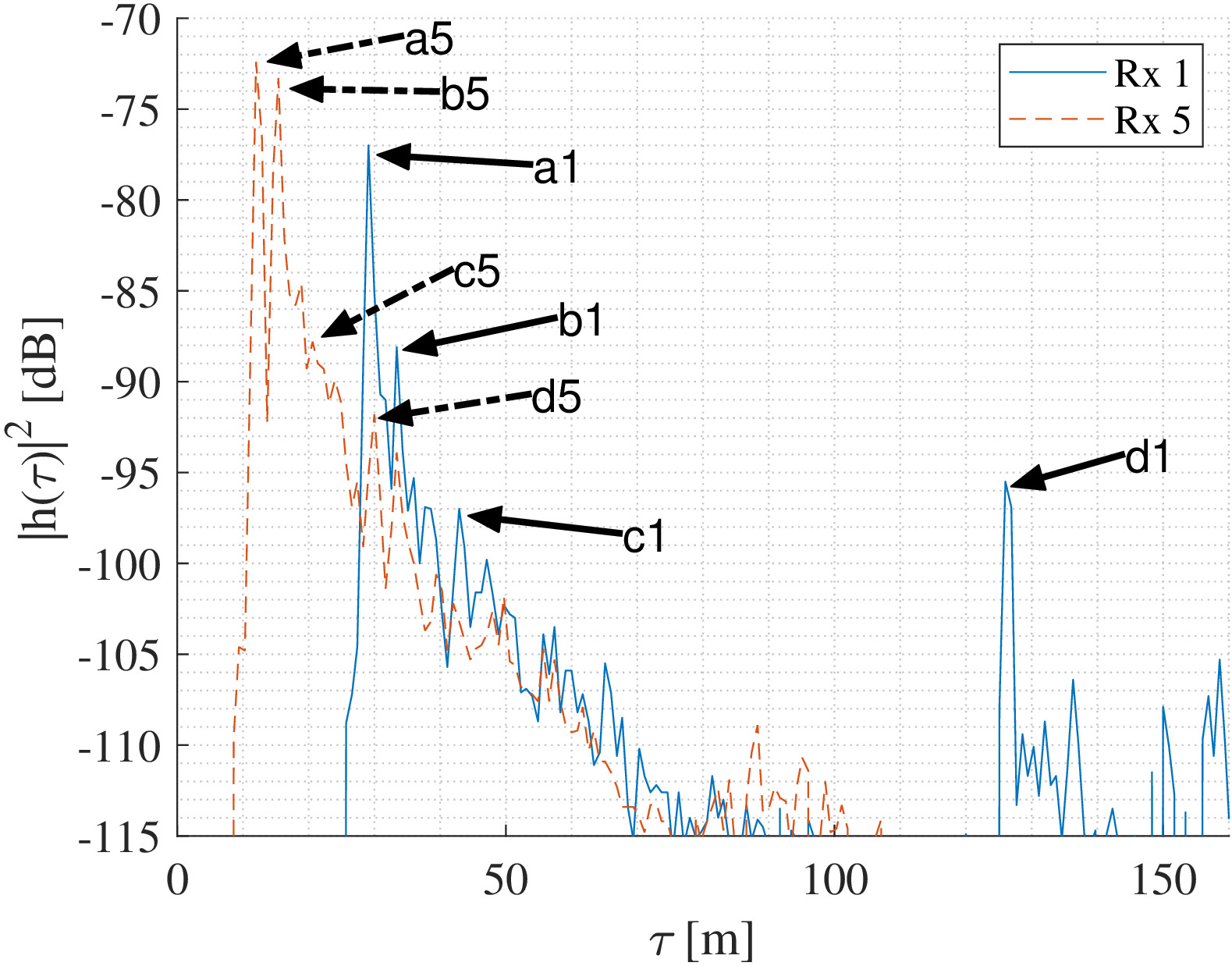}%
                    \label{fig: LoS PDP 1 data}%
                    }\\
                \subfloat[The geometry map for the short LoS PDP]{%
                    \includegraphics[width=1\linewidth, angle=0]{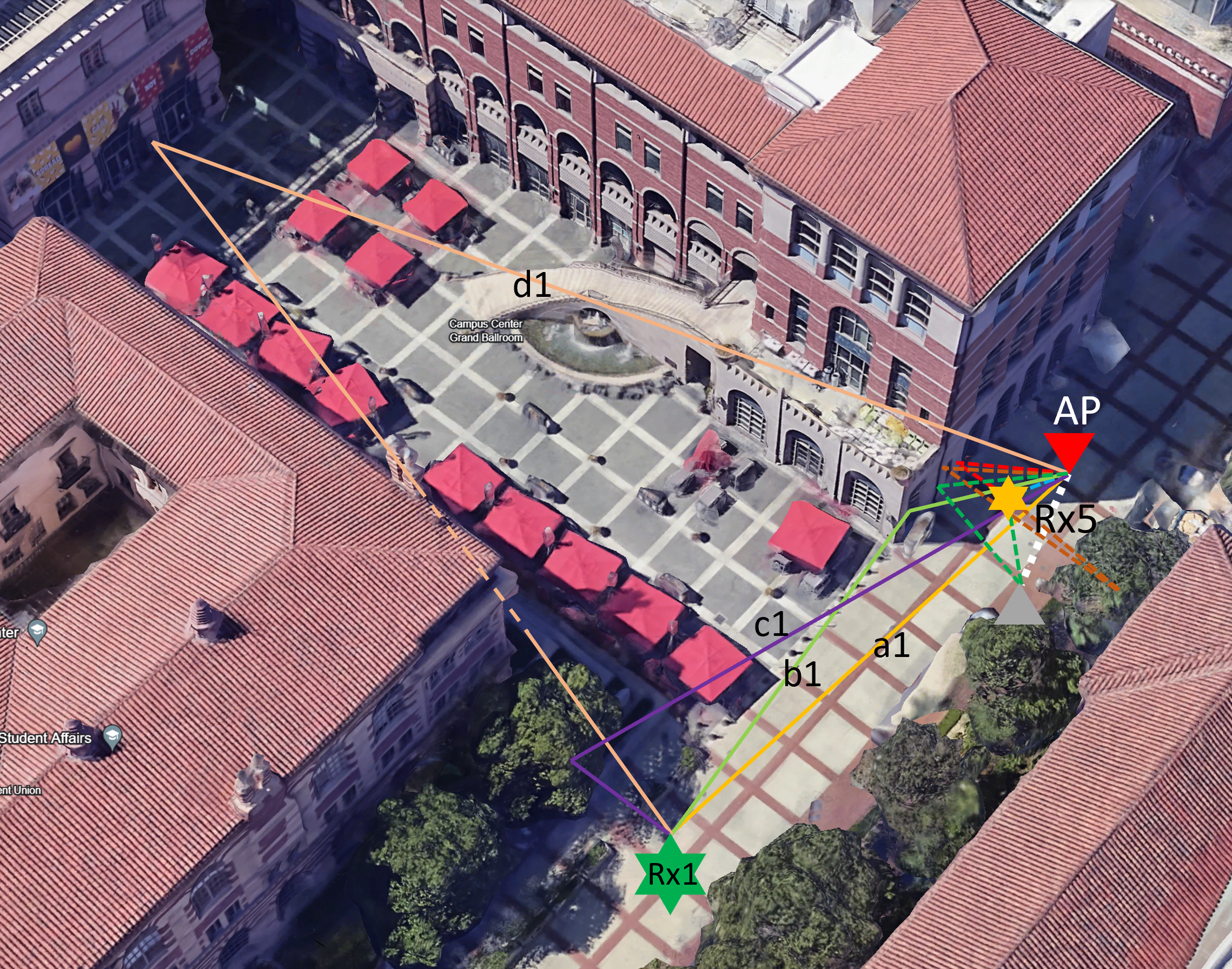}%
                    \label{fig: LoS PDP 1 map}%
                    }%
                \caption{LoS PDP sample where the AP is closer to the UE antennas. In b), the blue, red, green, and brown dash lines are paths $a5$, $b5$, $c5$, and $d5$.}
                \label{fig: LoS PDP 1}
            \end{figure}

            Even in situations with relatively short distances between \ac{Tx} and \ac{Rx}, the \ac{PDP} can exhibit surprising complexity, in particular in environments with inhomogeneous building structures. Fig. \ref{fig: LoS PDP 1 map} shows an environment in which two \acp{UE}, namely Rx1 and Rx5 of the \ac{UE} cluster $2$ are placed at a location where a street opens into a plaza. The red triangle indicates the \ac{AP} (raised into the air by the cherry picker), and the light gray triangle indicates the projection of the \ac{AP} on the ground (the white dashed line between the two thus is a line perpendicular to the ground), while the green and orange hexagrams are locations of \ac{Rx}1 and \ac{Rx}5, respectively. From the geometry, we can conjecture various possible propagation paths, which are labeled with alphanumeric acronyms. Table III lists the potential paths and their corresponding delays, as well as the delays of peaks in the \ac{PDP} that have similar delays. We can see that for Rx$5$, the \ac{AP} and \ac{UE} are very close, and not many propagation paths are feasible (most of those propagate ``almost vertically''). Some of the conjectured paths are ``nonspecular'', by which we mean that the direction of incidence and direction of reflection do not follow Snell's law, or that the plane of reflection cannot be determined (as occurs, e.g., for foliage, but also for the cherry-picker, which has many metal parts pointed into different directions), see the PDP in Fig. \ref{fig: LoS PDP 1 data}. Apart from the marked strong MPCs, the rest of the \ac{PDP} shows a single-exponential decay. For Rx1, the multipath structure is richer, though the difference between the first and the second peak in the \ac{PDP} is significantly larger. Most notable is that the \ac{PDP} exhibits a second cluster, associated with \acp{MPC} that are reflected at the back of the plaza, and thus provide a strong reflection (note that on the day of the measurement, the umbrellas visible in the photo were closed). Such multiple clusters lead to a significant increase in both \ac{DS} and Q-win.    
    
            \begin{table}[h]
                \centering
                \caption{Potential propagation MPCs and the corresponding routes for the short LoS sample}
                \begin{tabular}{|c|>{\centering\arraybackslash}p{1.2cm}|>{\centering\arraybackslash}p{1.1cm}|p{3.9cm}|}
                    \hline
                    \textbf{Path} & \textbf{Nominal Length} & \textbf{Actual Length} & \textbf{Explanation} \\ \hline \hline
                    a1 & 30.1 & 29.14 & Direct path \\ \hline
                    b1 & 35 & 33.43 & Building reflection/diffraction \\ \hline
                    c1 & 42.7 & 42.86 & Tree reflection (\textbf{non-specular}) \\ \hline
                    d1 & 126.5 & 126 & Building single reflection \\ \hline \hline
                    a5 & 13 & 12 & Direct path \\ \hline
                    b5 & 14.8 & 15.43 & Building single reflection \\ \hline
                    c5 & 22 & 21.43 & Building and the cherry picker double reflection (\textbf{non-specular}) \\ \hline
                    d5 & 30.5 & 30 & Building/foliage double reflection (\textbf{non-specular}) \\ 
                    \hline
                \end{tabular}
                \label{table: LoS PDP 1}
            \end{table}

        \subsubsection{NLoS PDP}
\begin{figure}[htp] 
                \centering
                \subfloat[NLoS PDP]{%
                    \includegraphics[width=1\linewidth, angle=0]{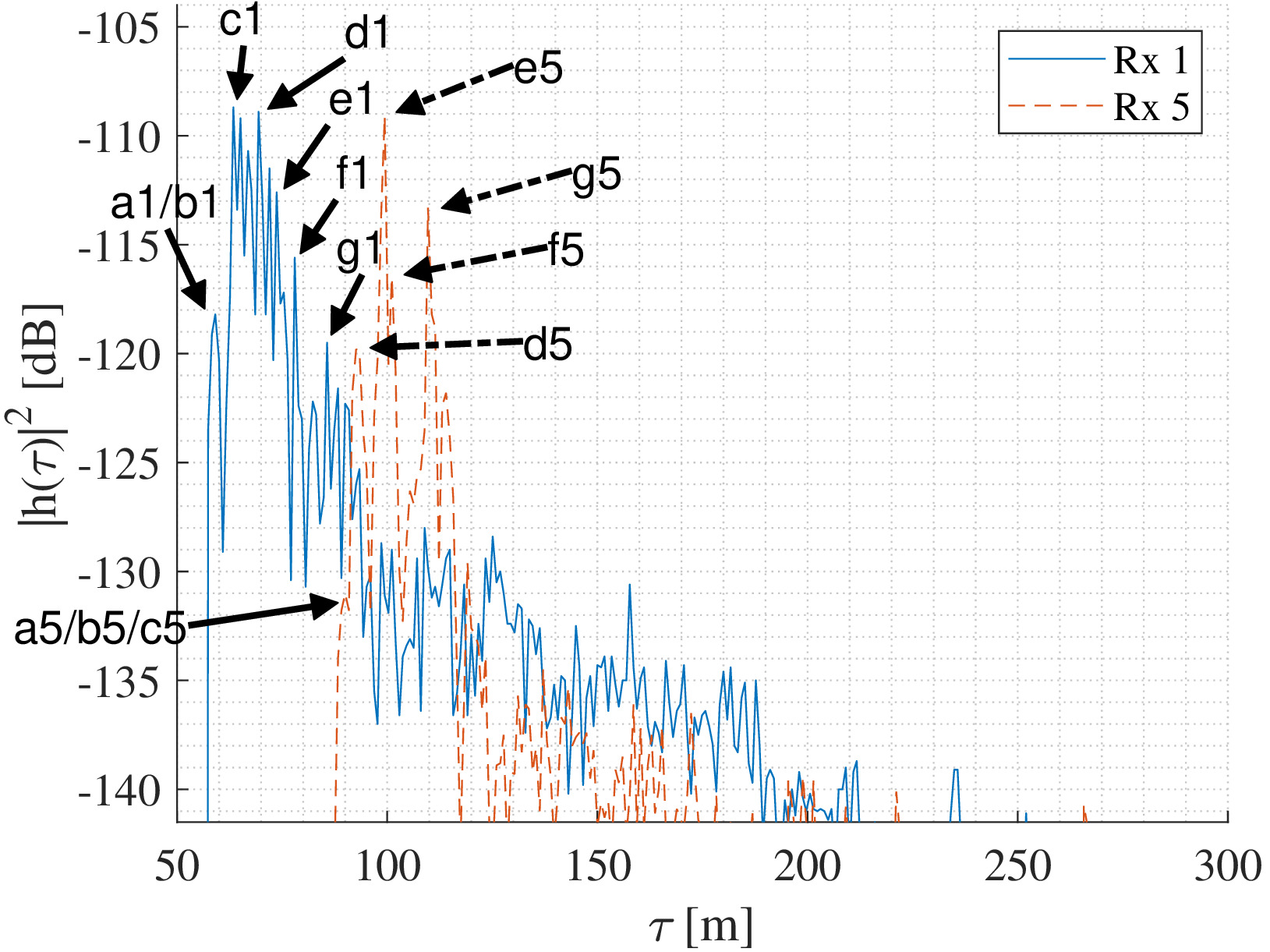}%
                    \label{fig: NLoS PDP data}%
                    } \\
                \subfloat[The geometry map for the NLoS PDP]{%
                    \includegraphics[width=1\linewidth, angle=0]{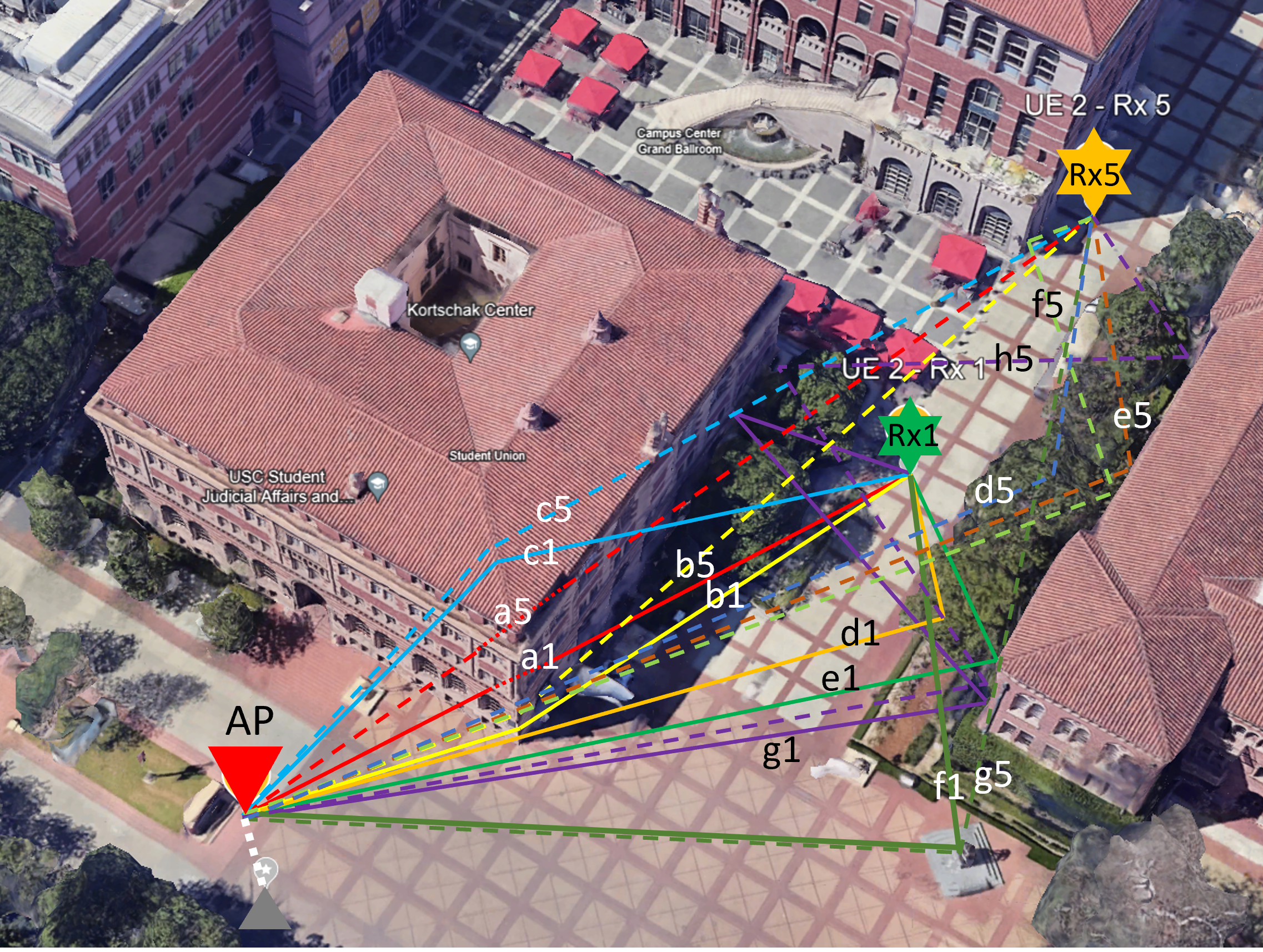}%
                    \label{fig: NLoS PDP map}%
                    }%
                \caption{NLoS PDP sample around a corner of a street}
                \label{fig: NLoS PDP}
            \end{figure}

            Both the relative importance of propagation processes, and the shape of the \ac{PDP}, show considerable differences to the \ac{LOS} case. Direct propagation through the building can occur when the distance traversed through it is short, though the resulting \ac{MPC} is considerably weaker than later-arriving components, leading to a ``soft onset'' of the \ac{PDP}. Diffraction around the building, and over the rooftop, are both important factors, due to the shallow diffraction angle. In Fig. \ref{fig: NLoS PDP data}, path $a1$/$b1$ is observable. However, it cannot be distinguished whether it is the in-building propagation, the horizontal diffraction around the square building, or the superposition of both because the two different paths share almost the same delay. Similarly, $a5$, $b5$, and $c5$ cannot be distinguished solely based on delay information. However, since the power at $90$ m is weak, it can be concluded for all three possible paths, the propagation losses are all high. One can also observe the reflection introduced by building walls and dense foliage by paths $d1$, $e1$, $g1$, $d5$, $e5$, $f5$, and $h5$. Particularly, we noticed a strong reflection from a copper statue at the lower right corner of Fig. \ref{fig: NLoS PDP map}. The reason for $g5$ having lower path loss than $f1$ could be the shape of the statue, which provides a higher reflection cross section area for Rx$5$ compared to the one in direction of Rx$1$. 
            
            There are also a large number of \acp{MPC} at large delays, though we could not identify the specific propagation processes, as there are too many possible paths with multiple interactions that could explain them. When the separation between \ac{Tx} and \ac{Rx} is even larger, a smaller number of \acp{MPC} carry significant power, all caused by single-reflection processes. Table \ref{table: NLoS PDP} summarizes the paths. 
        
            \begin{table}[h]
                \centering
                \caption{Potential propagation MPCs and the corresponding routes for the NLoS sample}
                \begin{tabular}{|c|>{\centering\arraybackslash}p{1.2cm}|>{\centering\arraybackslash}p{1.1cm}|p{3.9cm}|}
                    \hline
                    \textbf{Path} & \textbf{Nominal Length} & \textbf{Actual Length} & \textbf{Explanation} \\ \hline \hline
                    a1 & 60.7 & 60.9 & Direct link (through a building) \\ \hline
                    b1 & 60.9 & 60.9 & Building edge diffraction \\ \hline
                    c1 & 65.1 & 65.1 & Rooftop diffraction \\ \hline
                    d1 & 71.6 & 71.1 & Foliage single reflection (\textbf{non-specular}) \\ \hline
                    e1 & 75.3 & 75.4 & Building single reflection \\ \hline
                    f1 & 81 & 79.7 & Trojan Tommy reflection \\ \hline
                    g1 & 87.6 & 87.4 & Building double reflection \\ \hline \hline
                    a5 & 90.8 & 90 & Direct link (through building) \\ \hline
                    b5 & 91.5 & 90 & Building edge diffraction \\ \hline
                    c5 & 91.1 & 90 & Rooftop diffraction \\ \hline
                    d5 & 95 & 95.1 & Foliage single reflection (\textbf{non-specular}) \\ \hline
                    e5 & 102.6 & 102 & Building single reflection \\ \hline
                    f5 & 105.4 & 103.7 & Building double reflection \\ \hline
                    g5 & 110.3 & 112.3 & Trojan Tommy reflection \\ \hline
                    h5 & 134.8 & NaN & Building triple reflection \\
                    \hline
                \end{tabular}
                \label{table: NLoS PDP}
            \end{table}

    \subsection{Evolution of PDP}   \label{sec: AP-location-dependent PDP}
        A unique feature of our measurements is that we measure on trajectories of \acp{AP}. It is thus desirable to analyze the evolution of the measured \acp{PDP} with locations (indicated by time) of the \ac{AP}. This allows us to determine non-stationarities of the \acp{PDP} and have further sanity checks on the measurement results. These evolutions will also play a critical role in future CUNEC-type modeling for delay dispersion. Note that for space reasons, we discuss here only a smaller number of \acp{MPC} compared to the PDP snapshots.

\begin{figure}[htp] 
            \centering
            \subfloat[\ac{AP-LD} \ac{PDP} 1, the \emph{Time} axis indicates the AP locations. ]{%
                \includegraphics[width=1\linewidth, angle=0]{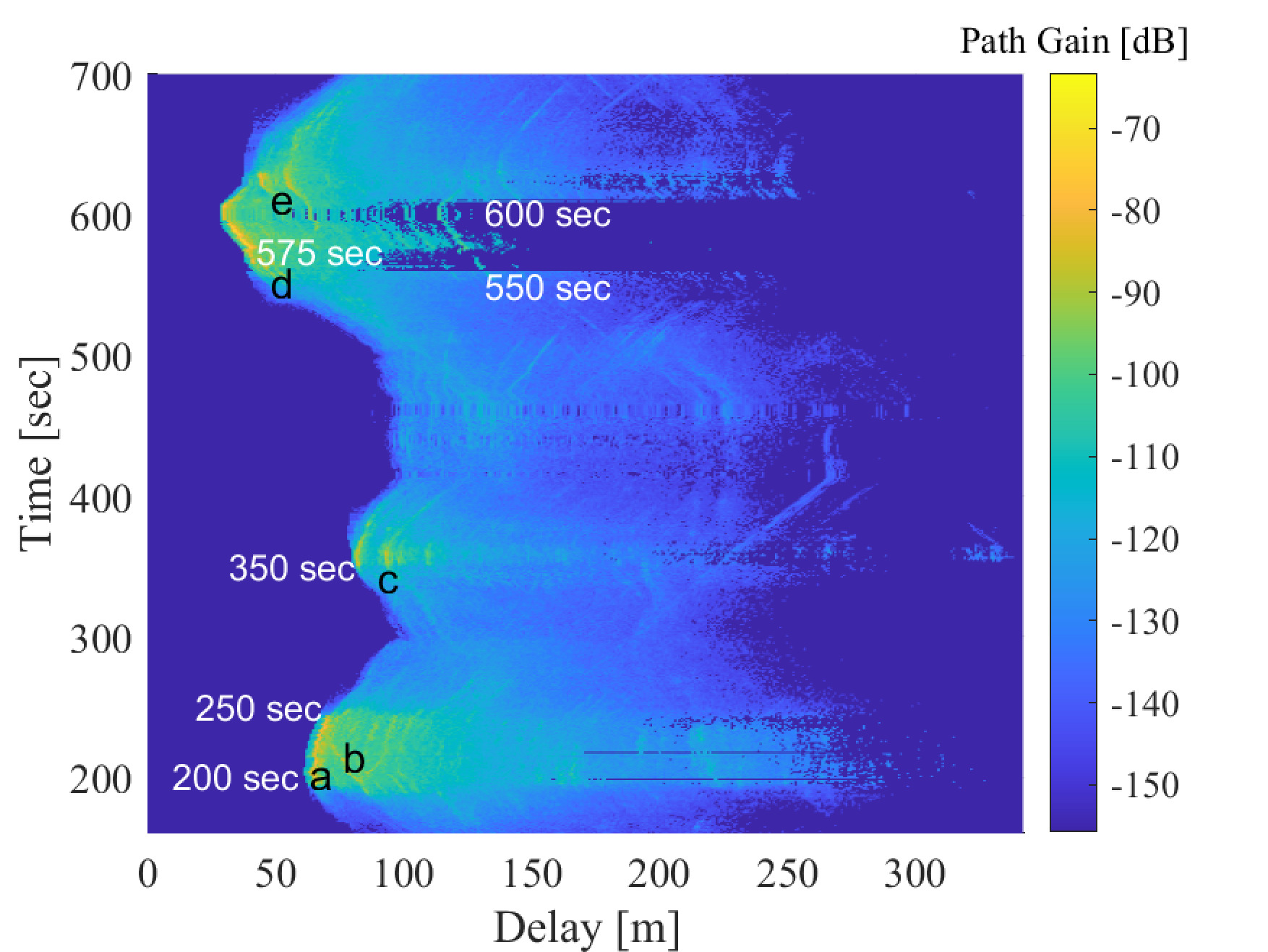}%
                \label{fig: AP-location-dependent PDP 1}%
                }\\
            \subfloat[Measurement route for \ac{AP-LD} \ac{PDP} 1. The green part of the trajectory is LOS, and the blue part is NLOS. Letters show the propagation effects giving rise to correspondingly labeled peaks in Fig. \ref{fig: AP-location-dependent PDP 1}.]{%
                \includegraphics[width=1\linewidth, angle=0]{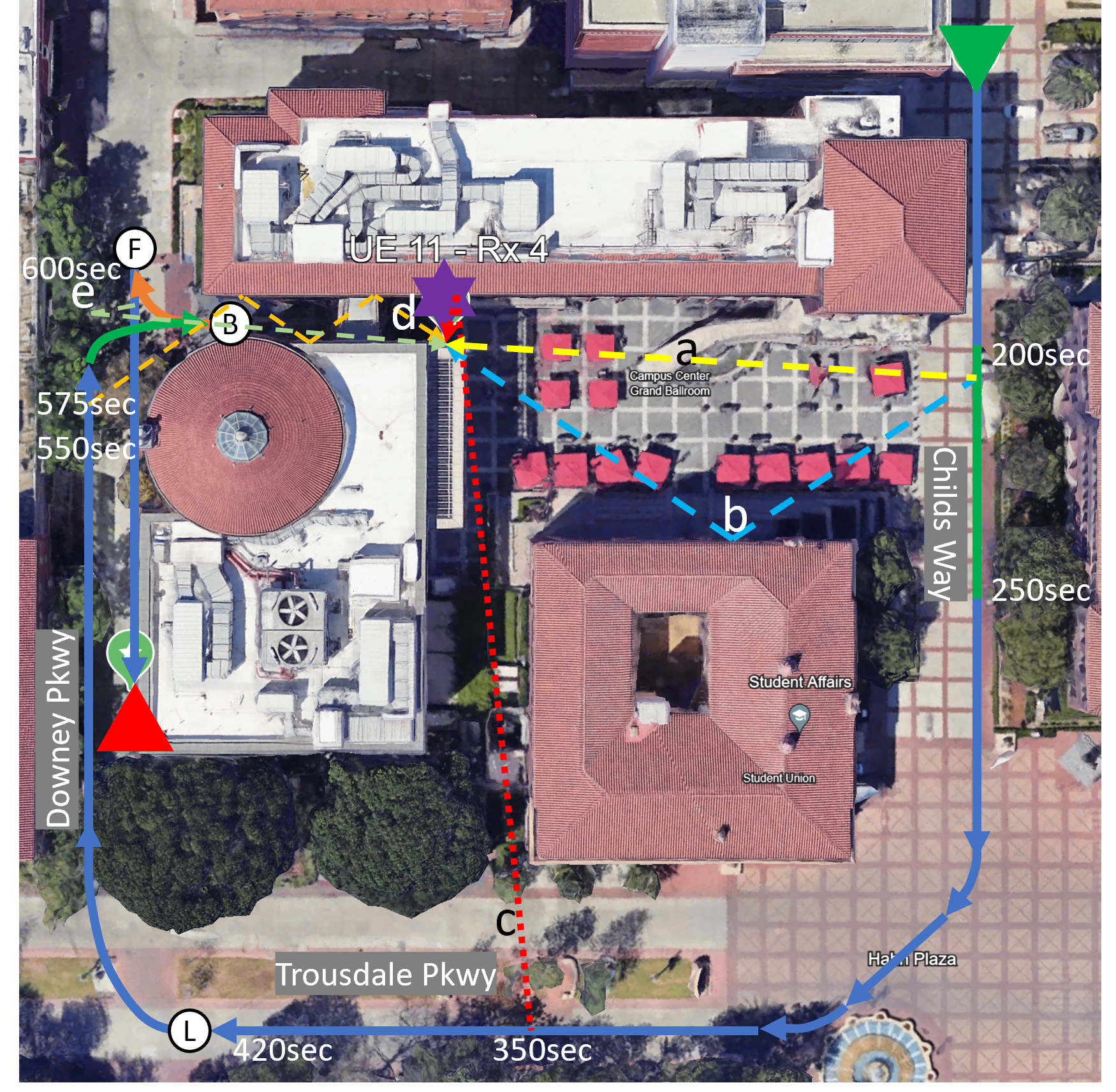}%
                \label{fig: AP-location-dependent PDP route 1}%
                }%
            \caption{\ac{AP-LD} \ac{PDP} at UE site 11, Rx 4}
            \label{fig: AP-location-dependent PDP example 1}
        \end{figure}

        The sample \ac{AP-LD} \ac{PDP} we discuss here is captured via a route shown in Fig. \ref{fig: AP-location-dependent PDP route 1}, between the AP and Rx$4$ in the \ac{UE} site $11$. The color representation is now different from Fig. \ref{fig: measurement locations}: the green part of the route now represents the LoS section, and the blue part is NLoS. The AP starts to move from the point at the upper right corner to the lower left of the map. The corresponding \ac{AP-LD} \ac{PDP} is shown in Fig. \ref{fig: AP-location-dependent PDP 1}. Due to the measurement principle of forming the virtual array, different \ac{AP} locations map to different time, i.e., the \emph{Time} axis in Fig. \ref{fig: AP-location-dependent PDP 1}. Rich scattering, i.e., many \acp{MPC}, can be observed throughout the trajectory. The first \ac{LOS} situation occurs when the AP moves downward on the map and is in front of the plaza (similar to the AP location in Fig. 5). The \ac{LOS} $a$ is the strongest component in this part of the \ac{PDP}, lasting from approximate $t=200$ to $t=250$, and its delay increases as the \ac{AP} moves further downwards. During the same time, the delay of \acp{MPC} $b$, which is reflected at the square building, decreases as the \ac{AP} moves downwards, as the excess path length via this building becomes shorter. However, this component dies out sooner than the \ac{LOS}, namely when the reflection on the building cannot fulfill Snell's laws anymore. For the subsequent $100$s, there are no pronounced \acp{MPC} that could be ascribed to a particular propagation process, though there is still significant energy and an analysis of PDP snapshots (not shown here) reveals an essentially monotonic exponential decay. Around $350$s, when the AP moves along the bottom horizontal trajectory, a gap between the building on the left and the Student Affairs building allows a low-attenuation channel, which is marked by $c$. While this is not strictly speaking a \ac{LOS}, a reflection of the long rectangular building at the top of the map provides an efficient propagation path. When the AP approaches the bottom left point marked by a circled letter ``L'', the cherry picker lowers the AP to $3$ m to avoid tree canopies when it goes into the ``1st U-turn'' street (the right half of Downey Way). Near $575$s, the \ac{AP} has \ac{LOS} to the \ac{UE} through the street canyon between the square buildings on the top and the left. Before it reaches this point, there are already multi-reflection components visible, such as $d$. The cherry picker then makes a U-turn at the location of the \ac{LOS} visibility. The two circled letters ``B'' and ``F'' represent the driving mode of the cherry picker, which stands for ``backward'' and ``forward'' (the orange route represents the reverse route of the cherry picker), such that the orientation of the array changes. Note that (for the part of the trajectory on the left side of the map) the \ac{ULA} is always facing to the right when the cherry picker is moving forward. Thus, when the cherry picker finishes the backward process, the \ac{ULA} now faces to the left, which makes $e$ happen: the \ac{LOS} component is very weak, but the reflection of the opposing house wall is stronger. 

        Due to the variations of the noise figure mentioned in Sec. \ref{sec: channel soudner}, the noise floor is higher between time $550$ sec and $600$ sec, resulting in shorter support of the \ac{PDP} after the noise thresholding. 

\begin{figure}[htp] 
            \centering
            \includegraphics[width=.9\linewidth]{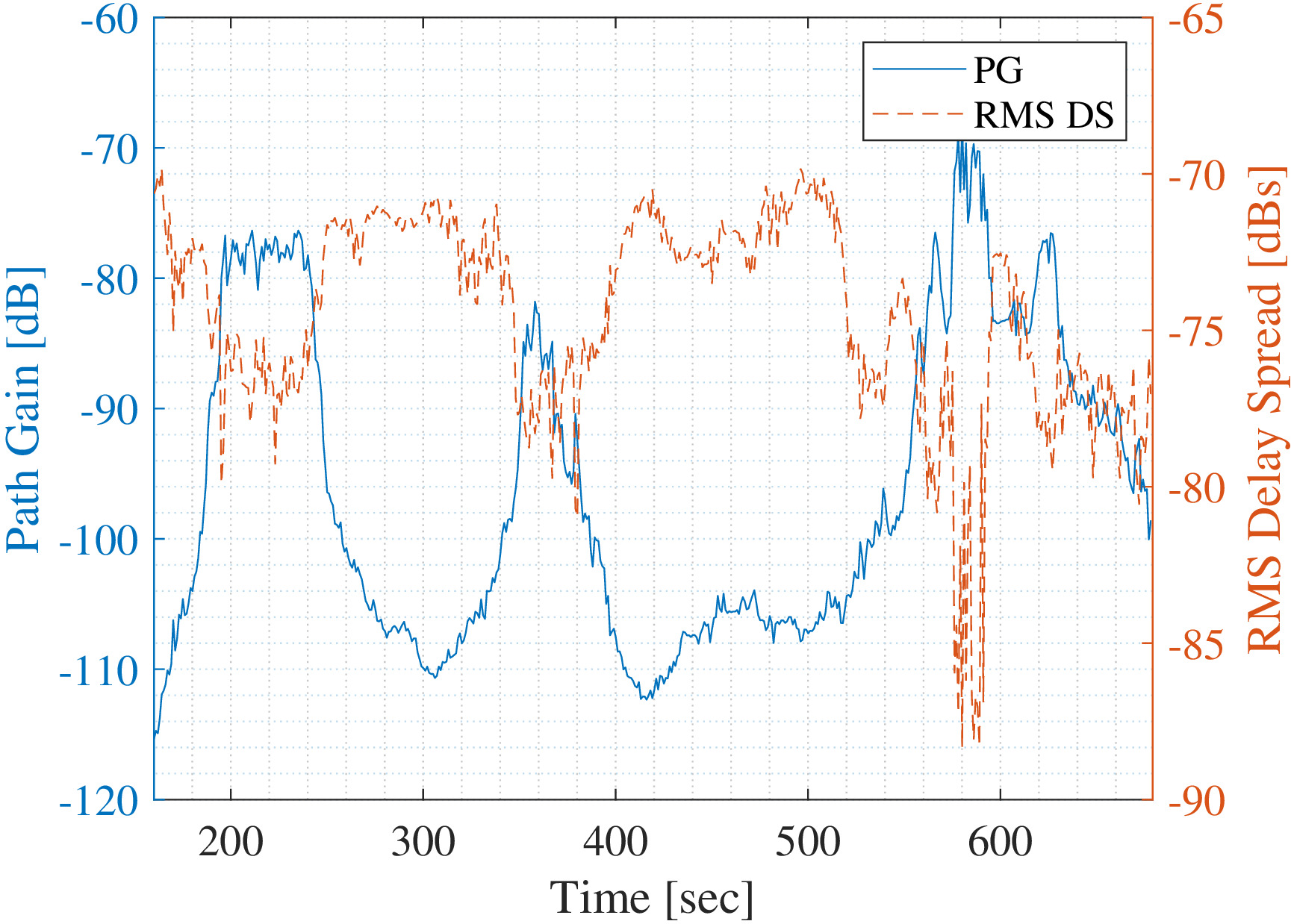}
            \caption{Channel parameters evolving with different \ac{AP} locations. }
            \label{fig: para vs AP location 1}
        \end{figure}

        Fig. \ref{fig: para vs AP location 1} shows the pathloss and \ac{RMS} \ac{DS} evolution for the same scenario. While there is a general trend that high path gain and low \ac{DS} occur together, there may be a significant deviation from this trend, e.g., at the end of the trajectory, where both path gain and \ac{DS} decrease. This may be related to the location of the \ac{UE}, i.e., between two close buildings. Such a narrow tunnel naturally filters out the \acp{MPC} with inappropriate incident directions (assuming specular reflections on building walls), and only the ones that share a similar propagation path from a certain direction can arrive at the \ac{UE}, which makes the \ac{DS} smaller. In the presence of \ac{LOS}, the pathloss varies mostly within $5$ dB when the distance does not change much, and the \ac{DS} mostly is below $-75$ dBs. \ac{NLOS} channels generally have both a higher \ac{DS} and lower path gain, and a typically negative correlation between the two parameters. 

        Other examples of AP-location \acp{PDP} are provided in \cite{zhang2024large}.

\section{Parameter Statistics} \label{sec: parameter statistics}
    We first describe the pathloss as a function of Euclidean distance. It must be kept in mind that this pathloss encompasses not only the actual attenuation by the channel, but is also influenced by the antenna pattern of both the \ac{Tx} and \ac{Rx}. This quantity is, by itself, meaningful when used for deployments with similar antennas; since patch antennas at the \acp{AP} are common, and antennas with dipole-like radiation characteristics are also used frequently for \acp{UE}, indeed deployments ``seeing'' similar pathloss characteristics can be expected. However, the impact of the antenna patterns must be kept in mind when interpreting the results. 

    \subsection{PL and Shadowing}
\begin{figure}[htp] 
            \centering
            \subfloat[LOS]{%
                \includegraphics[width=1\linewidth, angle=0]{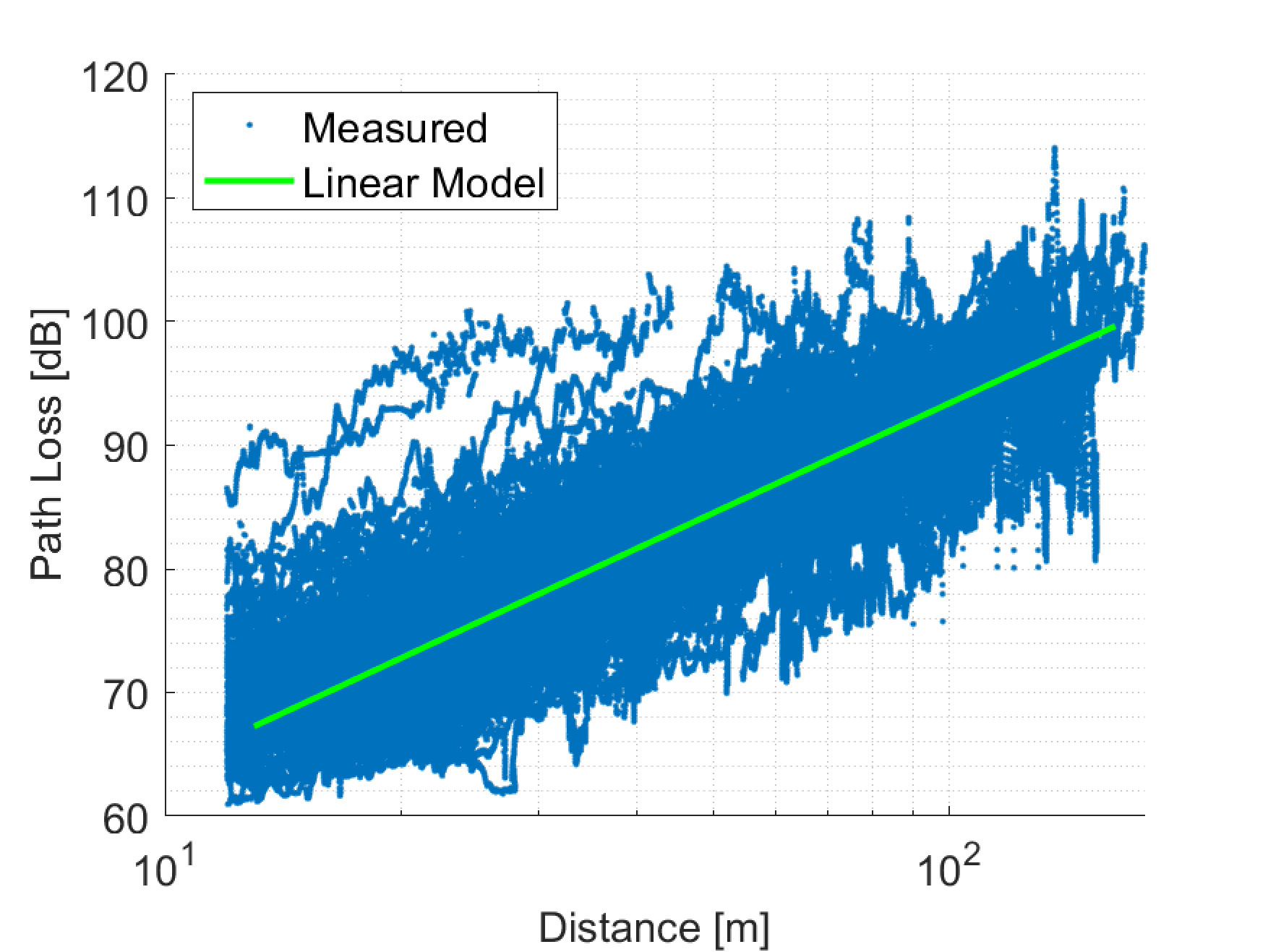}%
                \label{fig: LoS PL vs distance}%
                } \\
            \subfloat[NLOS]{%
                \includegraphics[width=1\linewidth, angle=0]{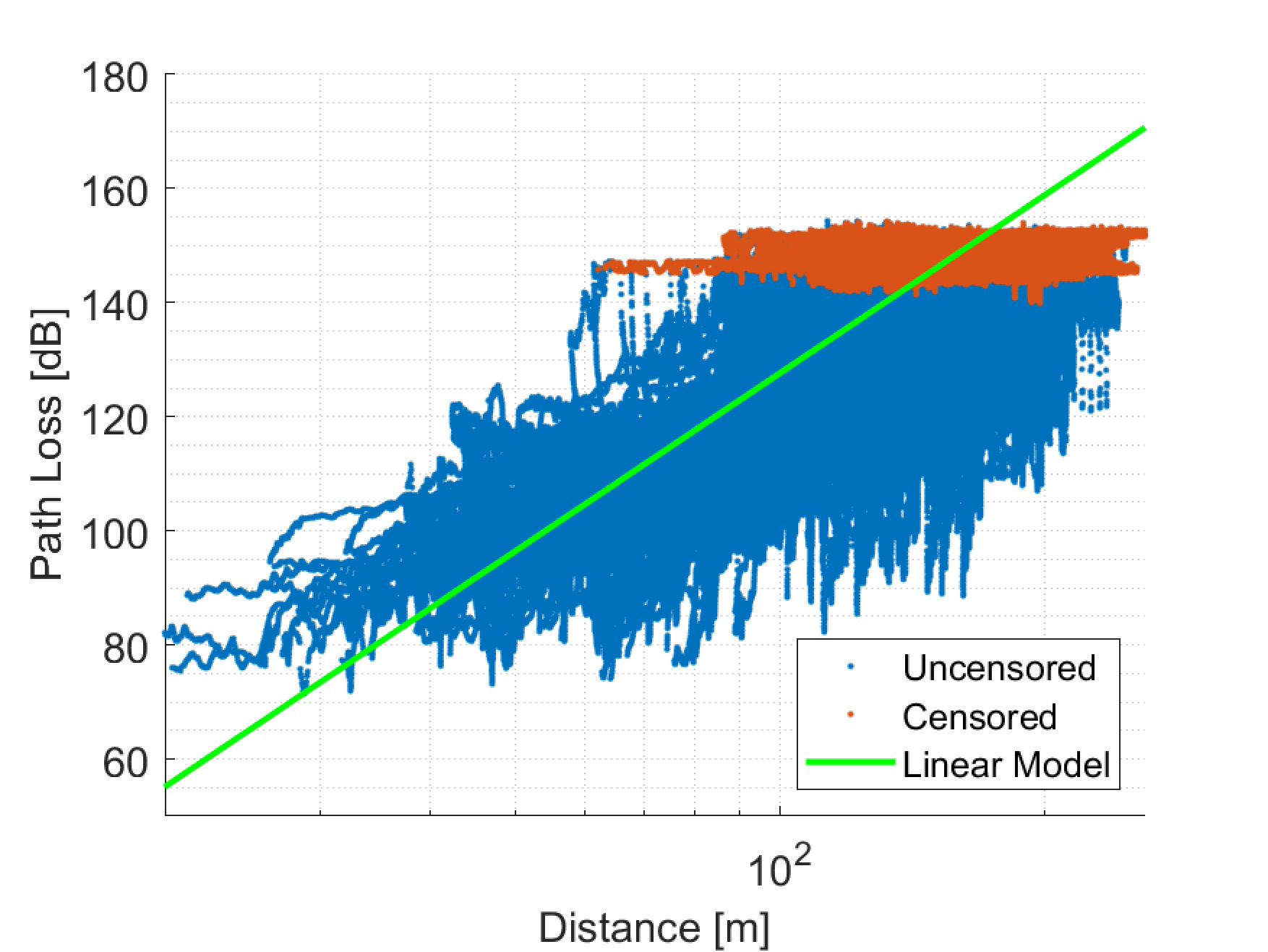}%
                \label{fig: NLoS PL vs distance}%
                }%
            \caption{Pathloss with linear model using logarithm distance. Bin-averaged with 5-m width. }
            \label{fig: PL vs distance}
        \end{figure}

    \setlength{\extrarowheight}{5pt}
        \begin{table*}
            \centering
            \caption{Pathloss and shadowing model parameters with $95\%$ confidence interval.}
            \label{tab: PL and shadowing stat}
            \begin{tabular}{|c|c|c|c|c|c|c|c|c|c|}
                \hline
                \multicolumn{1}{|c|}{\multirow{2}{*}{\textbf{Parameter}}} & \multicolumn{9}{c|}{\textbf{Parameters estimated with 95\% CI}} \\ \cline{2-10} 
                \multicolumn{1}{|l|}{} & \multicolumn{1}{c|}{$\alpha$} & \multicolumn{1}{c|}{$\alpha_{min,95\%}$} & \multicolumn{1}{c|}{$\alpha_{max,95\%}$} & \multicolumn{1}{c|}{$\beta$} & \multicolumn{1}{c|}{$\beta_{min,95\%}$} & \multicolumn{1}{c|}{$\beta_{max,95\%}$} & \multicolumn{1}{c|}{$\sigma_S$} & \multicolumn{1}{c|}{$\sigma_{S, min,95\%}$} & \multicolumn{1}{c|}{$\sigma_{S, max,95\%}$}\\ \hline \hline
                $PL^{\rm LoS}$ & \LOSPLAlpha & \LOSPLAlphaMin & \LOSPLAlphaMax & \LOSPLBeta & \LOSPLBetaMin & \LOSPLBetaMax & \LOSPLS & \LOSPLSMin & \LOSPLSMax \\ \hline
                $PL^{\rm NLoS}$ & \NLOSPLAlpha & \NLOSPLAlphaMin & \NLOSPLAlphaMax & \NLOSPLBeta & \NLOSPLBetaMin & \NLOSPLBetaMax & \NLOSPLS & \NLOSPLSMin & \NLOSPLSMax \\ \hline
            \end{tabular}%
        \end{table*}
        A plot of the pathloss as a function of 3-D Euclidean distance for the ensemble of all UE-AP combinations is shown in Fig. \ref{fig: LoS PL vs distance} and \ref{fig: NLoS PL vs distance}. Table \ref{tab: PL and shadowing stat} shows the fitting parameters using the $\alpha-\beta$ model and the shadowing standard deviation. The distance ranges of the underlying measurements, and thus the validity range for the model, are $[12, 178]$ m and $[20, 262]$ m, for \ac{LOS} and \ac{NLOS}, respectively. We firstly note that the \ac{LOS} pathloss shows a significantly higher standard deviation (around $\LOSPLSOneDigit$ dB) than is common in \ac{LOS} scenarios. This is partly due to the large number of measurement points taken in a variety of streets that have different structures (surface structure of the buildings, number of plazas and alleys intersecting them, etc.). Furthermore, these streets contain different amounts of foliage that might obstruct the \ac{LOS} \ac{MPC} as well as other \acp{MPC}. Finally, the fits are significantly impacted by the antenna patterns, as discussed above. This is true in particular for \ac{LOS} links, since in that case the pathloss will be heavily influenced by the antenna gains in the direction of the actual \ac{LOS} \ac{MPC}: when the \ac{AP} is close to the \ac{UE}, the \ac{LOS} direction may be at elevation angles as large as $170^\circ$ and thus - despite the mechanical downtilt of the \ac{AP} - antenna gains at both \ac{AP} and \ac{UE} might be very small, leading to a seeming increase in pathloss at small distances (we observed a similar effect in railway systems \cite{he2013measurement}). On the other hand, for large distances, the \ac{LOS} \ac{MPC} has a horizontally grazing angle of incidence at the \ac{AP} antenna, leading to some additional attenuation. Thus, while \ac{LOS} links in street canyons typically have $\alpha \le 2$, in our case we find $\alpha=\LOSPLAlpha$.  

        To compute the linear regression in the LOS plot, we use a binning of the measured pathloss points into bins of width $2$ m; this eliminates the impact of unequal number of measurement points at different distances (note that while the {\em spacing between} measurement points is essentially constant, the number of points at a given Euclidean distance from the \acp{UE} is different). Due to the height of the \ac{AP} of $13$ m and the \ac{UE} of $1$ m above ground, the minimum distance is $12$ m (we eliminate measurements during the time that the cherry picker is at the ground level or in the process of lifting the basket from the ground level upwards, as those points do not correspond to typical \ac{AP} deployments). 

        For the NLOS areas, there are locations where the received signal is censored (i.e., no signal power can be measured) because of the limitation of the system sensitivity.  Our system sensitivity is defined by the noise threshold, namely $\theta_{m',j}$. 
        Due to the nature of the dataset being the ensemble of all measurement campaigns, which spanned two weeks, with multiple system power cycles during transition and installation, a noise floor variation with a standard deviation of $\AveSTDNF$ dB can be observed. We perform the model fitting in the presence of censored data according to the method outlined in \cite{gustafson2015statistical} and \cite{karttunen2016path}.

        For the \ac{NLOS} case, we find that the pathloss coefficient is  than fir \ac{LOS}, namely around $\NLOSPLAlphaOneDigit$. This leads to a mean pathloss around $\LOSPLfittingValA$ vs $\NLOSPLfittingValA$ dB at  $\PLfittingDisA$ m and $\LOSPLfittingValB$ vs $\NLOSPLfittingValB$ dB at $\PLfittingDisB$ m distance, for \ac{LOS} and \ac{NLOS}, respectively. Higher attenuations exist, especially at even larger distances, but could not be measured due to the limited dynamic range of our setup.

    \subsection{Delay Spread}
\begin{figure}
            \centering
            \includegraphics[width=.9\linewidth]{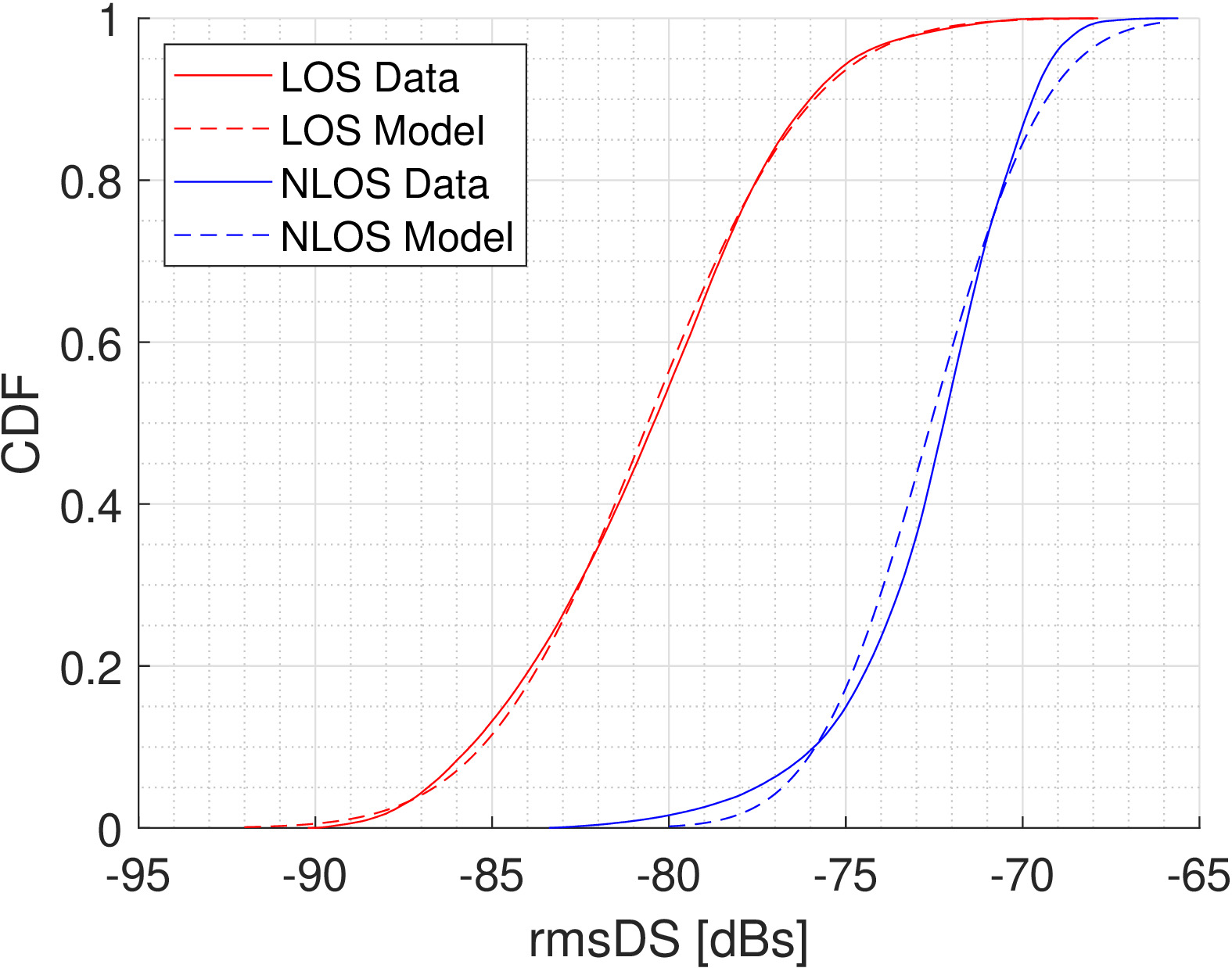}
            \caption{\ac{RMS} \ac{DS} model that fits Gaussian distribution in the logarithm scale. For LOS: $R^2=\LOSDSRsquare$ dB; for NLOS: $R^2=\NLOSDSRsquare$ dB. }
            \label{fig: RMS delay spread model}
        \end{figure}

        For the \ac{DS}, we only consider those locations where at least $20$ dB dynamic range is available in order not to bias the results. Taking the \ac{CDF} of the \ac{DS} separately for \ac{LOS} and \ac{NLOS} situations, we obtain Fig. \ref{fig: RMS delay spread model} to show the measured distribution in our \ac{LOS} and \ac{NLOS} scenarios.  

       The distribution of the \ac{RMS} \ac{DS} is commonly modeled as a log-normal distribution, i.e., the \ac{DS} in dBs is described by a normal distribution. We can see that the fits are good, with $R^2$ value of $\LOSDSRsquare$ and $\NLOSDSRsquare$, respectively\footnote{From the empirical data, a Log-Normal fit to the \ac{DS} values on a dB scale provides a slightly better fit with parameter $R^2=\NLOSDSRsquareLogNormal$ dB for \ac{NLOS}.}.
      The fitting parameters are shown in Table \ref{tab: RMS delay spread stat}; median \ac{DS} are on the order of $\LOSDSMedian$ and $\NLOSDSMedian$ dBs, respectively. While this is small for an urban canyon environment, the microcellular setup (i.e., antennas below rooftop height) and thus the absence of far reflections explain the behavior. 

        \setlength{\extrarowheight}{5pt}
        \begin{table*}
            \centering
            \caption{\ac{RMS} \ac{DS} model parameters with $95\%$ confidence interval.}
            \label{tab: RMS delay spread stat}
            \begin{tabular}{|c|c|c|c|c|c|c|}
                \hline
                \multicolumn{1}{|c|}{\multirow{2}{*}{\textbf{Parameter}}} & \multicolumn{6}{c|}{\textbf{Statistical model parameters estimated with 95\% CI}} \\ \cline{2-7} 
                \multicolumn{1}{|l|}{} & \multicolumn{1}{c|}{$\mu$} & \multicolumn{1}{c|}{$\mu_{min,95\%}$} & \multicolumn{1}{c|}{$\mu_{max,95\%}$} & \multicolumn{1}{c|}{$\sigma$} & \multicolumn{1}{c|}{$\sigma_{min,95\%}$} & \multicolumn{1}{c|}{$\sigma_{max,95\%}$} \\ \hline \hline
                $\sigma_\tau^{\rm LoS}$ & \LOSDSMu & \LOSDSMuMin & \LOSDSMuMax & \LOSDSSigma & \LOSDSSigmaMin & \LOSDSSigmaMax \\ \hline
                $\sigma_\tau^{\rm NLoS}$ & \NLOSDSMu & \NLOSDSMuMin & \NLOSDSMuMax & \NLOSDSSigma & \NLOSDSSigmaMin & \NLOSDSSigmaMax \\ \hline
            \end{tabular}%
        \end{table*}

\begin{figure}
            \centering
            \subfloat[LOS]{%
                \includegraphics[width=1\linewidth, angle=0]{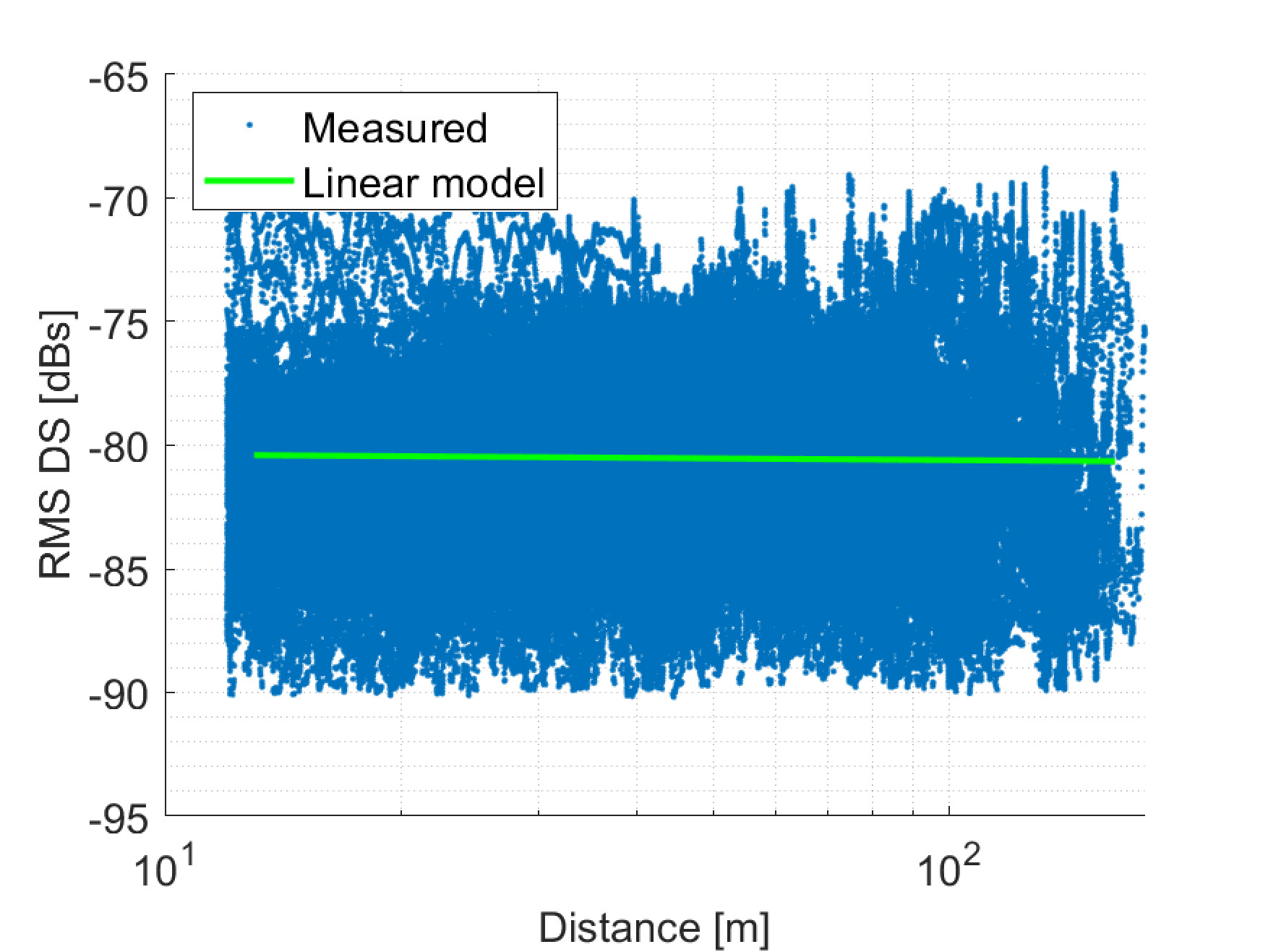}%
                \label{fig: LoS RMSDS vs distance}%
                } \\
            \subfloat[NLOS]{%
                \includegraphics[width=1\linewidth, angle=0]{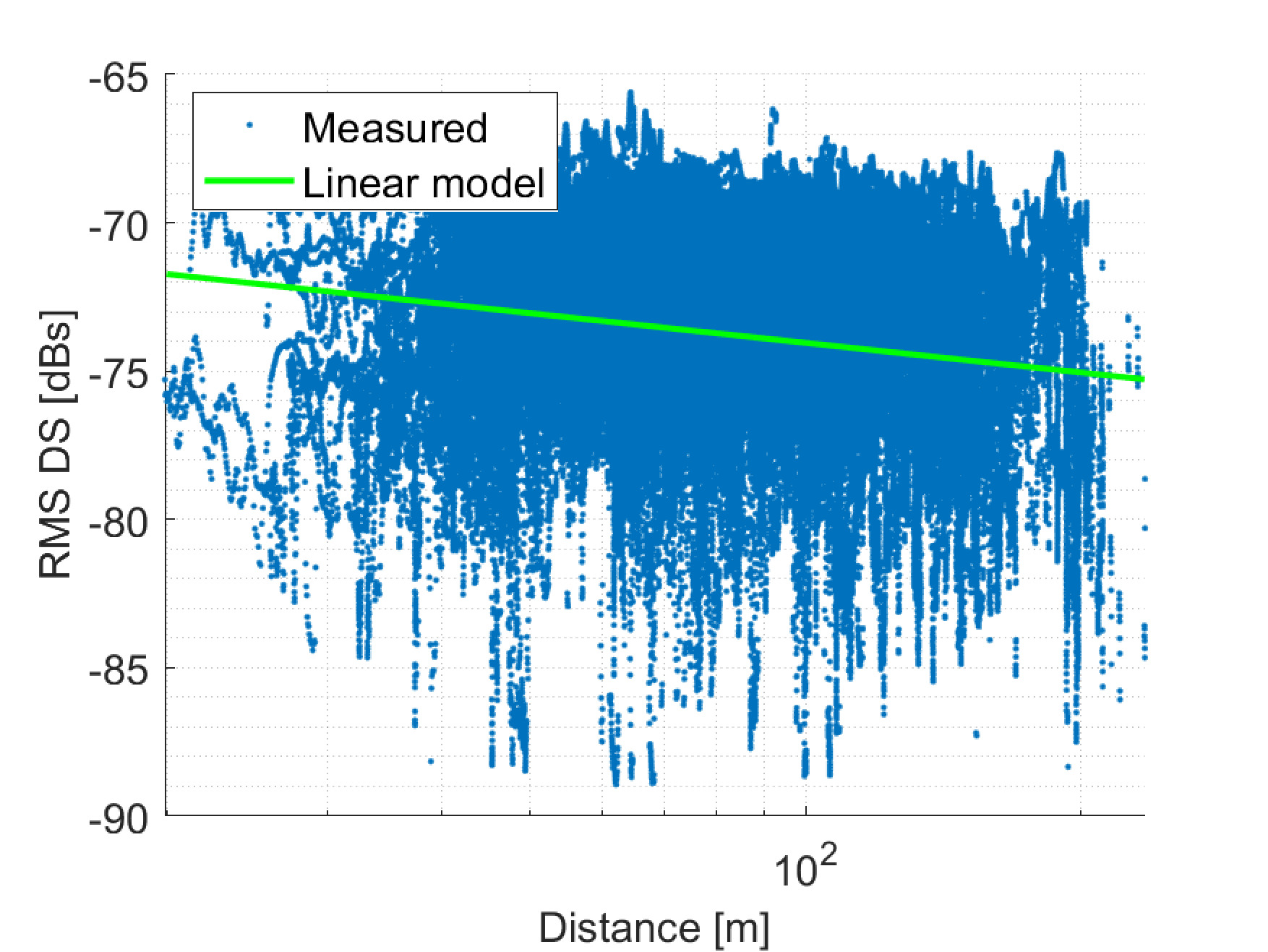}%
                \label{fig: NLoS RMSDS vs distance}%
                }%
            \caption{RMS DS with linear model using logarithm distance. Bin-averaged with 5-m width. }
            \label{fig: RMSDS vs distance}
        \end{figure}

        Furthermore, we evaluated a linear fit of the \ac{DS} as a function of \ac{AP}-\ac{UE} distance. Fig. \ref{fig: LoS RMSDS vs distance} and \ref{fig: NLoS RMSDS vs distance} show the results for \ac{LOS} and \ac{NLOS} scenarios, respectively. We found the \ac{DS} is almost independent of the distance (slope is $\LOSDSAlpha$) for \ac{LOS}. 
        We conjecture that this is due to the fact that the \ac{DS} is mostly dominated by the immediate surroundings, such as the vicinity to plazas and street intersections, not the actual distance between \ac{AP} and \ac{UE}.
        As for the \ac{NLOS} case, a negative slope of $\NLOSDSAlpha$ occurs. This is somewhat in contrast to traditional models that describe the delay spread as increasing with distance. Again, we conjecture that this can be explained by the  (sometimes significantly) below-rooftop position of the \ac{AP} prevents the occurrence of long-delayed components that normally increase the \ac{DS} for larger distances, and restricts the contributing \acp{MPC} to, essentially, waveguiding in the street canyon, and where the runlength differences between the guided components decrease as the distance increases.

    \subsection{Window parameters}
\begin{figure}
            \centering
            \includegraphics[width=.9\linewidth]{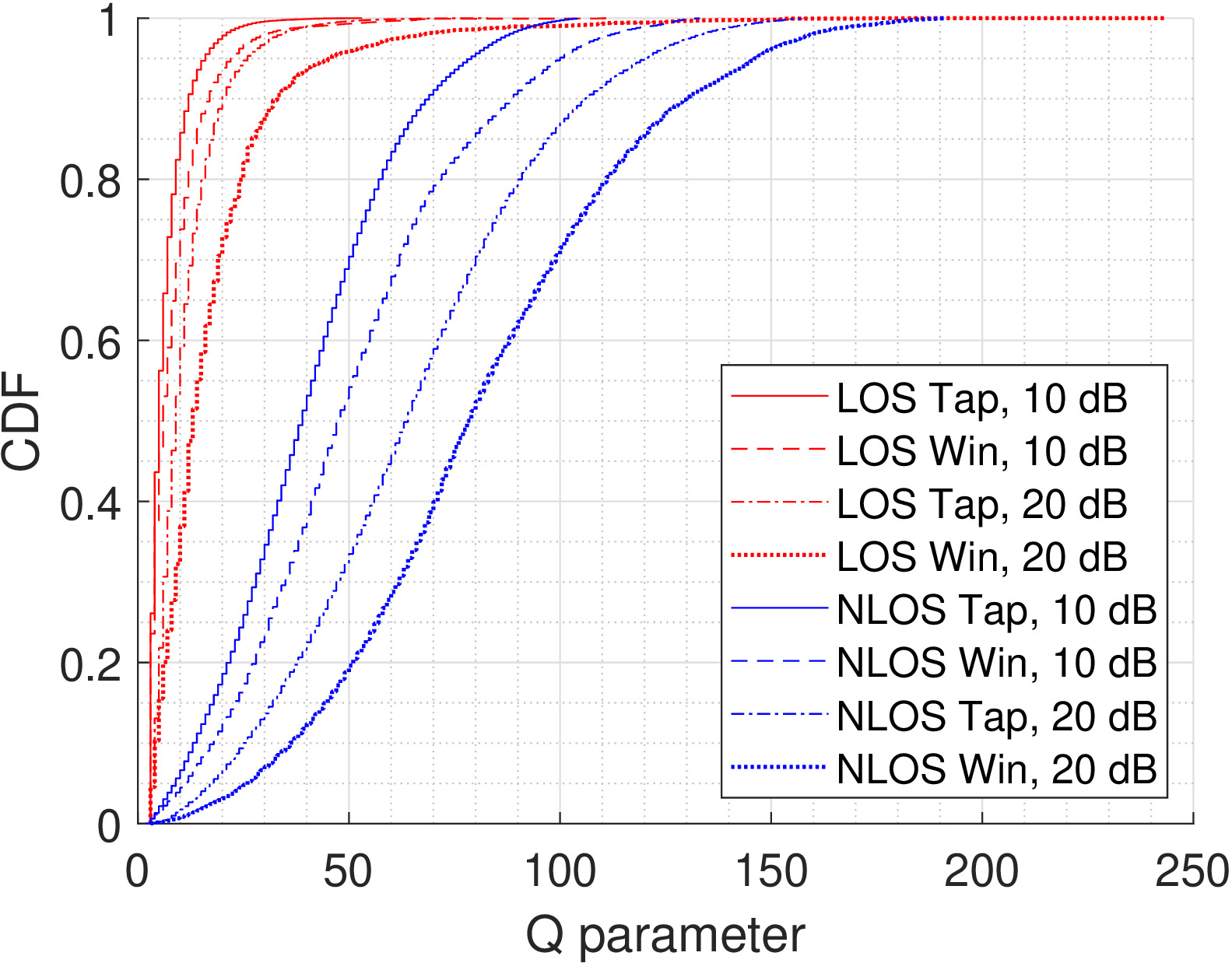}
            \caption{Power distribution using quotient parameters, Q-tap and Q-window, with different required \acp{SIR}}
            \label{fig: Qpara}
        \end{figure}

        Lastly, we analyze the distributions of the Q-win and Q-tap parameters, as defined and interpreted in Sec. \ref{sec: parameter processing}. They are expressed in units of resolvable delay bins (while for rectangular spectrum, this is usually assumed to be the inverse \ac{BW}; the Kaiser filtering broadens it by a factor of 1.2). Naturally, these distributions show significant differences for \ac{LOS} and \ac{NLOS} situations, as shown in Fig. \ref{fig: Qpara}. Furthermore, they depend on the target SIR and the desired outage level (in the following, assumed to be $10$\%). For $10$ dB, the Q-tap parameters are $\LOSQtap$ and $\NLOSQtap$ and the Q-win parameters are $\LOSQwin$ and $\NLOSQwin$ taps, for \ac{LOS} and \ac{NLOS} situations, respectively. The fact that Q-win and Q-tap differ from each other indicates that the PDPs show some sparsity. Non-negligible energy contributions at long-delayed \acp{MPC} need to be considered when designing the equalizers, rather than clustering closely. 

\section{Conclusions}
    We have performed an extensive measurement campaign for determining the pathloss and delay dispersion of wideband \ac{CF-mMIMO} channels. The dataset encompasses channels between more than $20,000$ \ac{AP} locations and $\NumUETotal$ \ac{UE} locations, which is orders of magnitude larger than previous multi-BS outdoor measurements. After describing the measurement setup that is based on a combination of virtual and switched array, we presented sample results that explain the peaks in the measured \acp{PDP} in terms of the surrounding geometry and the specific propagation processes leading to each peak. Observations of the evolution of the \acp{MPC} from the plot of the \ac{PDP} over time further confirm the measurement results and show the significant variations with the time/location of the \acp{AP}. Finally, we extracted the statistics of the ``classical'' channel models, namely slope and intercept of the $\alpha - \beta$ model, shadowing standard deviation, \ac{DS}, Q-win, and Q-tap. 

    The sample results give insights not only into the propagation channels, but also how \ac{CF-mMIMO} systems would perform with different densities of \acp{AP}, and number of \acp{AP} belonging to one cluster. The differences in pathloss between \ac{LOS} and \ac{NLOS} are significant, namely up $50$ dB (for the same distance AP-), indicating that deployment should be dense enough to guarantee that each \ac{UE} has \ac{LOS} to at least one \ac{AP}. Obstruction of the \ac{LOS} by vegetation is also a significant factor. At the same time, the strong difference between \ac{LOS} and \ac{NLOS} indicates that interference from nodes that have \ac{NLOS} might be limited, which might enable smaller \ac{AP} clusters connected to each \ac{UE}. It is also remarkable that not only the channel transfer function, but also the \ac{PDP}, and thus the frequency correlation function, can differ significantly for the links from one \ac{UE} to different \acp{AP}, which might complicate acquisition and exploitation of second-order channel statistics. 
    
    Finally, the traditional channel statistics such as $\alpha - \beta$ pathloss model and \ac{DS} are valuable for system simulations when the underlying generic channel model is based on those specific quantities, as is the case in most theoretical investigations of \ac{CF-mMIMO} systems. However, the experimental results in this paper confirm that there are significant variations between statistics of different streets, and that the long-term correlations must be taken into account on a per-street basis. Future work will thus focus on the development of a wideband channel model that can properly quantify these effects, generalizing our CUNEC approach that we previously introduced for the modeling of pathloss.

\section{Acknowledgment}
The authors thank the USC WiDeS group members for their contributions, in particular Hussein Hammoud, Bowei Wu, Ashwani Pradhan, Kelvin Arana, Pramod Krishna, Tianyi Yang, Tyler Chen, Ishita Vasishtha, Haoyu Xie, and Linyu Sun, for their help in system calibration, measurement campaign, and data evaluation.

\bibliographystyle{IEEEtran}
\bibliography{reference.bib}
\end{document}